\documentclass[%
 reprint,
 amsmath,amssymb,
 aps,
]{revtex4-2}

\usepackage[mode=buildnew]{standalone}
\usepackage{subcaption}
\usepackage{color}
\usepackage{graphicx}
\usepackage{dcolumn}
\usepackage{bm}
\usepackage{float}
\usepackage{braket}

\usepackage[hidelinks]{hyperref}
\def\equationautorefname#1#2\null{%
  Eq.#1(#2\null)%
}
\def\tableautorefname#1#2\null{%
  {table#1#2\null}%
}
\def\sectionautorefname#1#2\null{%
  {section#1#2\null}%
}
\def\figureautorefname#1#2\null{%
  {Fig.#1#2\null}%
}
\def\subsectionautorefname#1#2\null{%
  {section#1#2\null}%
}

\usepackage{tikz}
\usetikzlibrary{automata,arrows,positioning,calc,fit}

\begin{document}

\preprint{APS/123-QED}

\title{Operationally Accessible Uncertainty Relations for Thermodynamically Consistent Semi-Markov Processes}

\author{Benjamin Ertel}
\author{Jann van der Meer}%
\author{Udo Seifert}%
\affiliation{%
 II. Institut für Theoretische Physik, Universität Stuttgart, 70550 Stuttgart, Germany
}%

\date{\today}


\begin{abstract}
Semi-Markov processes generalize Markov processes by adding temporal memory effects as expressed by a semi-Markov kernel. We recall the path weight for a semi-Markov trajectory and the fact that thermodynamic consistency in equilibrium imposes a crucial condition called direction-time independence for which we present an alternative derivation. We prove a thermodynamic uncertainty relation that formally resembles the one for a discrete-time Markov process. The result relates the entropy production of the semi-Markov process to mean and variance of steady-state currents. We prove a further thermodynamic uncertainty relation valid for semi-Markov descriptions of coarse-grained Markov processes that emerge by grouping states together. A violation of this inequality can be used as an inference tool to conclude that a given semi-Markov process cannot result from coarse-graining an underlying Markov one. We illustrate these results with representative examples.
\end{abstract}

\maketitle


\captionsetup{justification=raggedright}

\section{Introduction}
Semi-Markov processes are a class of stochastic processes with various applications in physics. They serve as a mathematical description for continuous-time random walks \cite{montroll1965,Tunaley1974,Tunaley1975,Tunaley1976,Kutner2017_Rev,Masuda2017_Rev} and can also be used to describe particular cases of coarse-grained Markov processes \cite{Qian2007,martinez2019,Teza2020}. Continuous-time random walks can model different physical processes, especially anomalous diffusion
\cite{klafter1987,bouchaud1990, meerschaert2014,Barkai2014_Rev}. Related statistics are observed in miscellaneous experiments, including electronic transport in nanocrystalline electrodes \cite{Nelson1999} and diffusion processes in porous media \cite{Koch1988,Cortis2006,Ferreira2016} and biophysical systems \cite{Hoefling2013_Rev,Metzler_Exp2011,Goiko2018}. The theory of semi-Markov processes is also used in other fields aside from physics. Semi-Markov processes are discussed in queuing and renewal theory and can be applied to describe processes in finance and economy \cite{Fabens1961,Korolyuk1975,janssen2013,Scalas_2006}. 

For Markov processes that describe physical or (bio)chemical processes in contact with one or several heat baths, a comprehensive framework called stochastic thermodynamics has been developed \cite{Seifert2012}. Some of its concept and results have been extended to semi-Markov processes, in particular the crucial role time-reversibility of trajectories plays in equilibrium \cite{chari1994,Qian2007}, the identification of entropy production \cite{Girardin2003}, the derivation of a corresponding fluctuation theorem \cite{Esposito2008, andrieux2008} and the application of large deviation results \cite{Maes2009}. 


A prominent result for Markov processes is the thermodynamic uncertainty relation (TUR), a bound on current fluctuations in terms of the entropy production \cite{barato2015, gingrich2016}. Recent work includes numerous applications and refinements of this result as reviewed in \cite{seifert2019, Horowitz2020}. A somewhat technical generalization to semi-Markov process has been achieved in \cite{vu2020}. However, this TUR includes a memory term that is not operationally accessible by observing currents and their fluctuations.


This work aims at complementing the description of semi-Markov processes from the perspective of stochastic thermodynamics. We formulate an operationally accessible TUR valid for a general semi-Markov process. For semi-Markov processes that emerge by grouping states of an underlying Markov process, we derive an even stronger bound dubbed here the Markov-based TUR. Since this bound makes use of the underlying Markov network, it can be violated for a general semi-Markov process. Operationally, observing a violation of this bound then excludes the existence of an underlying Markov network.

This paper starts with a review of the fundamental concepts of semi-Markov processes from a thermodynamic perspective in \autoref{sec:2}. A TUR valid for general semi-Markov processes is formulated in \autoref{sec:3}. Semi-Markov processes emerging from coarse-graining a Markov network are discussed from the perspective of the Markov-based TUR in \autoref{sec:4}. Examples illustrating the results are presented in \autoref{sec:5}. Section VI contains a concluding perspective. Technical aspects are relegated to the appendix.

\section{Fundamentals of Semi-Markov Processes}\label{sec:2} 
\subsection{From Markov to semi-Markov}

The dynamics of many discrete physical systems, for example molecular motors or chemical reaction networks, can be described as Markov processes. Within this description, transitions from state $I$ to state $J$ are described by transition rates $k_{IJ}$ and a Poisson-type waiting time distribution $\psi_{I}(t) = \Gamma_{I}\exp(-\Gamma_{I}t)$ with an escape rate  $\Gamma_{I} = \sum_{J}k_{IJ}$ for each state $I$ of the network. A transition from $I$ to $J$ occurs with probability $p_{I\to J} = k_{IJ}/\Gamma_{I}$. Combining the distributions of waiting time $t$ and jump destination $J$ results in a kernel of the form
\begin{equation}
  \psi_{I\rightarrow J}\left(t\right) = k_{IJ}\exp\left(-\Gamma_{I}t\right)
  \label{Eq_Int:Markov}    
,\end{equation} i.e., the probability density of observing a jump from $I$ to $J$ after waiting in $I$ for a time period $t$. Based on \autoref{Eq_Int:Markov}, a path weight description and thermodynamic quantities defined on single trajectories can be identified \cite{Seifert2012,Seifert2005}.

In many physical systems however, the waiting time distribution $\psi_{I}(t)$ of observable states $I$ is not given by a Poisson distribution and the resulting dynamics is therefore not Markovian. An exemplary class of systems exhibiting this behavior are biophysical processes at the level of single cells with typically many degrees of freedom (see, e.g., \cite{Golding2006,Dix2008,Hoefling2013_Rev}). Thus, a correct description of these systems calls for a generalization of Markov processes to arbitrary waiting time distributions $\psi_{I}(t)$ or even to waiting processes $\psi_{I \to J}(t)$ that depend on the next destination $I \to J$. One possible class of generalized stochastic processes fulfilling this condition are semi-Markov processes.

\subsection{Definition}
Given a discrete set of states, a Markov process is fully characterized by the current state $I$ of the system. In a semi-Markov process, a full description has to include the waiting time $t$ that has passed since the last transition to the current state $I$ additionally.

The characteristic quantity of a state $I$ is its semi-Markov kernel $\psi_{I\rightarrow J}\left(t\right)$. Formally, the kernel is defined as the joint distribution of waiting time $t$ and jump destination $J$ if the actual state is $I$ with age zero, i.e.,
\begin{equation}
\psi_{I\rightarrow J}\left(t\right) = p\left(I_{n+1}=J; t_{n+1}-t_{n}=t|I_{n}=I\right)
\label{II:EQ1_Psi}
,\end{equation}
where $t_i$ denotes the time at which the $i$-th jump takes place from $I_i$ to $I_{i+1}$. Markov processes are semi-Markov processes with an exponential semi-Markov kernel as defined in \autoref{Eq_Int:Markov}.

Semi-Markov processes possess temporal memory because the waiting times in each state are needed for the full characterization of the process. The probability for surviving in state $I$ up to $t$ given the system has already survived up to $t_{0}$ generally depends on $t_{0}$ unless the system is Markovian. Markov processes do not possess temporal memory because the associated waiting time distribution is a Poisson distribution, which implies $\psi_{I\to J}(t_{0}+t)/\psi_{I\to J}(t_{0}) = \psi_{I\to J}(t)$.

\subsection{Construction of the path weight}
\begin{figure}[bt]
\includegraphics[width=\linewidth]{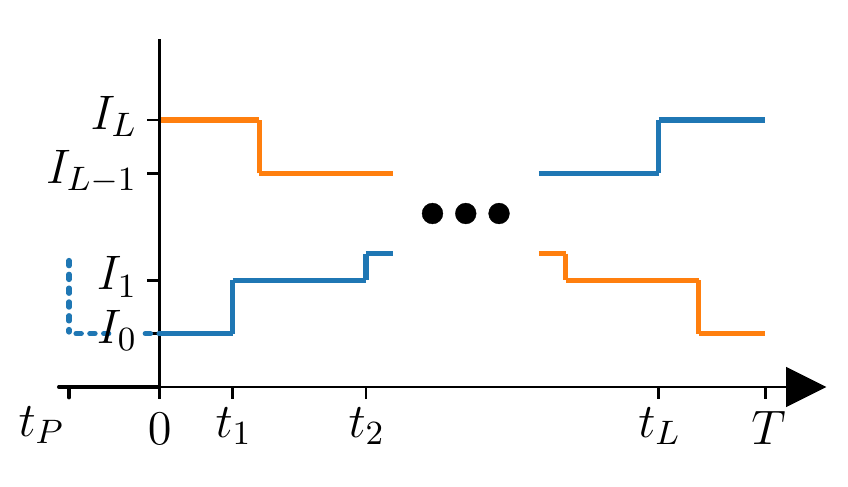}
\caption{Generic semi-Markov trajectory $\Gamma$ starting in $I_{0}$ with time-reversed counterpart $\tilde{\Gamma}$ starting in $I_{L}$. The time $t_{P}$ at which the system jumped into $I_{0}$ lies somewhere in the past.}
\label{II:Fig1_GenSM} 
\end{figure}

Semi-Markov processes can be described with a path weight quantifying the probability of a specific trajectory $\Gamma$. A general semi-Markov trajectory $\Gamma$, shown in \autoref{II:Fig1_GenSM} in dark color, starts in $I_{0}$ at time $0$ and ends up in $I_{L}$ at time $T$ with $L$ jumps at times $t_{i}$ in-between. Step $i$ includes waiting in state $I_{i}$ for the duration $t_{i+1} - t_{i}$ and jumping to state $I_{i+1}$ afterwards. The probability for this single step of waiting and jumping is $\psi_{I_{i}\rightarrow I_{i+1}}\left(t_{i+1}-t_{i}\right)$. Up to boundary terms, this observation leads to the path weight 
\begin{eqnarray}
    \mathcal{P}\left[\Gamma\right] &&= p_{I_{0}}\left(0\right)\Psi_{I_{0}\rightarrow I_{1}} \left(t_{1}\right)\nonumber\\ &&\times\prod_{i=1}^{L-1}\psi_{I_{i}\rightarrow I_{i+1}}\left(t_{i+1}-t_{i}\right)\Phi_{I_{L}}\left(T-t_{L}\right)
\label{II:EQ1_PW_Vor}  
\end{eqnarray}
for $\Gamma$, as the repeating pattern is identical for all intermediary states in the trajectory. This product corresponds to the probability for observing $L-1$ steps along the trajectory with the wait-and-jump pattern described above. 

The last factor in \autoref{II:EQ1_PW_Vor} is the survival probability in state $I_{L}$, which is given by
\begin{equation}
    \Phi_{I_{L}}\left(T-t_{L}\right) = \sum_{J}\int_{T-t_{L}}^{\infty}\psi_{I_{L}\rightarrow J}\left(t\right) dt.
\label{II:EQ2_Phi}    
\end{equation}
The system does not jump out of $I_{L}$ for at least $T-t_{L}$, which happens with the probability given by \autoref{II:EQ2_Phi}. The sum over all possible destinations of the jump $J$ is needed because the true destination of the jump is not known.

The first step has to be treated differently. The trajectory starts in $I_{0}$ at $t = 0$ with probability $p_{I_{0}}\left(0\right)$ given by the initial conditions of the process. The kernel for the first jump, $\Psi_{I_{0}\rightarrow I_{1}} \left(t_{1}\right)$, has to be distinguished from the other kernels because of the temporal memory of semi-Markov processes. As shown in \autoref{II:Fig1_GenSM}, $\Gamma$ enters state $I_{0}$ at $t_{P} < 0$ in the past. Because $t_{P}$ is not known in general, an appropriate average is needed.

\subsection{Past average}
The past average $\Psi_{I\rightarrow J}\left(t'\right)$ is equal to the probability for observing the partial waiting time $t'$ in state $I$ with jump destination $J$ given that the full, true waiting time $t_i$ is not known but was drawn from the semi-Markov kernel $\psi_{I\rightarrow J}\left(t\right)$.
The observed waiting time is always smaller than the full waiting time in the state. As the jump to $J$ is not relevant for the following derivation, the state indices can be dropped for simplicity and will be reinserted at the end.

We approximate a continuous waiting time distribution $\psi\left(t\right)$ to arbitrary precision with a sum of delta functions with weights $w_{i}$ at different times $t_{i}$ in the form
\begin{equation}
    \psi\left(t\right) = \sum_{i}w_{i}\cdot\delta\left(t-t_{i}\right).
\label{II:EQ3_Psi_Delta}  
\end{equation}
By using the law of total probability, the past average becomes
\begin{equation}
   \Psi\left(t'\right) = \sum_{i}p\left(t'|t_{i}\right)p\left(t_{i}\right)
\label{II:EQ4_P_Split} 
.\end{equation}
Here, $p\left(t'|t_{i}\right)$ denotes the probability for observing the partial waiting time $t'$ given that the full waiting time $t_{i}$ is drawn from $\psi\left(t\right)$, whereas $p(t_i)$ is the probability that the full waiting time is $t_{i}$.

The conditioned probability term can be expressed as
\begin{equation}
p\left(t'|t_{i}\right) = \frac{1}{t_{i}}\int_{0}^{t_{i}}\delta\left(t' - t\right)dt = \frac{1}{t_{i}}\Theta\left(t_{i} - t'\right)   
\label{II:EQ5_P_Hist}    
,\end{equation}
where $\Theta\left(t_{i} - t'\right)$ is the Heaviside step function, as $t'$ is uniformly distributed between $0$ and $t_i$.
 
Understanding the term $p(t_i)$ is somewhat subtle. Starting the observation of a given trajectory at a random time instant, say $t=0$, is a typical instance of the inspection paradox \cite{pal2018}. Since realizations of a larger waiting time have a higher probability to be observed, the probability that the actual waiting time of the observed step is $t_i$ is proportional to both the corresponding weight $w_i$ and the waiting time $t_i$ itself. This leads to
\begin{equation}
    p\left(t_{i}\right) = \frac{w_{i}t_{i}}{\sum_{i}w_{i}t_{i}} = \frac{w_{i}t_{i}}{\langle t\rangle},
\label{II:EQ6_P_Traj} 
\end{equation}
where $\langle t\rangle$ is the mean waiting time in the actual state.

Substituting \autoref{II:EQ5_P_Hist} and \autoref{II:EQ6_P_Traj} into \autoref{II:EQ4_P_Split} and taking the limit of infinitely many delta functions representing $\psi\left(t\right)$ leads to
\begin{equation}
    \Psi_{I\rightarrow J}\left(t'\right) = \frac{\int_{t'}^{\infty}\psi_{I\rightarrow J}\left(t\right)dt}{\langle t_{I}\rangle}
\label{II:EQ7_Past} 
\end{equation}
after reincluding the state indices of the jump process.

If the initial waiting time $t_{i}$ was known, taking the past average would not be necessary. However, this would require knowledge about the process for $t < 0$ in the past. As this knowledge is inaccessible for a trajectory defined on $0 \leq t \leq T$ by definition, the past average is crucial for the introduced path weight.

The expression in \autoref{II:EQ7_Past} for the past average is identical to the commonly used expression derived from the stationary solution of the dynamics \cite{Qian2007,Maes2009} and can also be derived from a true, unweighted average over all possible uncertain pasts \cite{Esposito2008}. 

The past average is redundant for Markov processes because of their missing temporal memory. This property can be verified after substituting \autoref{Eq_Int:Markov} with corresponding first moment $\langle t_{I}\rangle = 1/\Gamma_{I}$ into \autoref{II:EQ7_Past} to obtain $\Psi_{I\rightarrow J}\left(t'\right) = \psi_{I\rightarrow J}\left(t'\right)$.

\subsection{Thermodynamic consistency in equilibrium and direction-time independence}
Not all semi-Markov processes are thermodynamically consistent in equilibrium. Demanding time-reversibility of the path weight in equilibrium results in a mathematical condition on $\psi_{I\rightarrow J}\left(t\right)$ emerging from the boundary terms of the path weight. The probability of drawing a forward trajectory $\Gamma$ or its corresponding backward trajectory $\tilde{\Gamma}$ should be the same in equilibrium, which means that the path weights for both trajectories have to be identical.

The backward trajectory $\tilde{\Gamma}$ is constructed by time-reversal of $\Gamma$ leading to  $\tilde{\Gamma}\left(t\right) = \Gamma\left(T-t\right)$. The time-reversed trajectory for the generic case is shown in \autoref{II:Fig1_GenSM} in bright color. The path weight $ \mathcal{P}[\tilde{\Gamma}]$ can be constructed in the same way as for the forward trajectory, leading to
\begin{eqnarray}
    \mathcal{P}[\tilde{\Gamma}] &&= p_{I_{L}}\left(T\right)\Psi_{I_{L}\rightarrow I_{L-1}} \left(T-t_{L}\right)\nonumber\\ &&\cdot\prod_{i=1}^{L-1}\psi_{I_{i}\rightarrow I_{i-1}}\left(t_{i+1}-t_{i}\right)\Phi_{I_{0}}\left(t_{1}\right).
\label{II:EQ8_PW_Back}  
\end{eqnarray}
The path weights $\mathcal{P}[\Gamma]$ and $\mathcal{P}[\tilde{\Gamma}]$ are equal in equilibrium only for semi-Markov kernels $\psi_{I\rightarrow J}\left(t\right)$ which factorize into a product of two terms
\begin{equation}
    \psi_{I\rightarrow J}\left(t\right) = p_{I\rightarrow J}\cdot\psi_{I}\left(t\right)
\label{II:EQ9_DTI}
.\end{equation}
The probability for a jump from $I$ to $J$ is given by $p_{I\rightarrow J} = \int_{0}^{\infty}\psi_{I\rightarrow J}\left(t\right)dt$, which, in particular, does not depend on the waiting time $t$ in $I$. Similarly, $\psi_{I}\left(t\right) = \sum_{J}\psi_{I\rightarrow J}\left(t\right)$ is the distribution of waiting times in state $I$ regardless of the destination of the following jump. If the semi-Markov kernel $\psi_{I\rightarrow J}\left(t\right)$ is separable in the sense of \autoref{II:EQ9_DTI}, the waiting time in $I$ will not depend on the destination $J$ of the following jump. 

Equation (11), also called direction-time independence (DTI) condition, is a necessary condition for time-reversal symmetry of semi-Markov processes in equilibrium \cite{Qian2007,chari1994}. An intuitive derivation is given in Appendix A, whereas a rigorous treatment under slightly more general conditions is found in the original proof \cite{chari1994}. Following \cite{Esposito2008, andrieux2008, Maes2009}, we demand that the DTI condition is satisfied for processes both in and out of equilibrium.

\vspace{-0.1029cm}

\subsection{Entropy production and semi-Markov currents}
The total entropy production along a trajectory is a natural measure of irreversibility in the underlying stochastic process. The total entropy production along a semi-Markov trajectory is given by
\begin{eqnarray}
    \Delta S_{\text{tot}}\left[\Gamma\right] &&\equiv \ln\frac{\mathcal{P}[\Gamma]}{\mathcal{P}[\tilde{\Gamma}]}\nonumber\\
    &&= \ln\frac{p_{I_0}\left(0\right)}{p_{I_L}\left(T\right)} + \ln\frac{\langle t_{I_L}\rangle}{\langle t_{I_0}\rangle} + \sum_{i}\ln\frac{p_{i-1\rightarrow i}}{p_{i\rightarrow i-1}}\nonumber\\
    &&= \ln\frac{p_{I_0}\left(0\right)}{p_{I_L}\left(T\right)} + \sum_{i}\ln\frac{\langle t_{i}\rangle p_{i-1\rightarrow i}}{\langle t_{i-1}\rangle p_{i\rightarrow i-1}}
    \label{II:EQ10_S}
.\end{eqnarray}
The path weights from \autoref{II:EQ1_PW_Vor} and \autoref{II:EQ8_PW_Back} are used together with DTI from \autoref{II:EQ9_DTI} for the second line; the third line follows by extension of the fraction inside the sum.

This entropy production was introduced in \cite{Girardin2003} and used for example in \cite{Esposito2008,Maes2009}. Equation (12) generalizes the entropy production along a Markov trajectory \cite{Seifert2005}, which can be verified by substituting \autoref{Eq_Int:Markov} for the semi-Markov kernel, fixing transition probabilities and first moments as $p_{I\rightarrow J} = k_{IJ}/\Gamma_{I}$ and $\langle t_{I}\rangle = 1/\Gamma_{I}$, respectively.

The final sum in \autoref{II:EQ10_S} can be interpreted in terms of a stochastic current along a semi-Markov trajectory. The generalized stochastic current $\mathcal{J}$ along a trajectory $\Gamma$ of length $T$ is given by
\begin{equation}
    \mathcal{J}[\Gamma] = \frac{1}{T}\sum_{IJ}n_{I\rightarrow J}q_{IJ}  
\label{II:EQ11_JGen}
,\end{equation}
with $n_{I\rightarrow J}$ the total number of jumps from state $I$ to $J$ along the trajectory and $q_{IJ}$ an antisymmetric matrix of increments. We denote the mean and variance of the random variable $\mathcal{J}$ by $\langle\mathcal{J}\rangle$ and $\text{Var }\mathcal{J}$, respectively.

A mean probability current can be defined in terms of the stochastic description without referring to the particular trajectory $\Gamma$ \cite{Qian2007}. 
The dynamics of a semi-Markov process can be described by a generalized master equation from which the steady state $\pi$ can be calculated. In this steady-state, the probability current $j_{IJ}$ between state $I$ and $J$ is naively given by the difference of forward and backward flow
\begin{equation}
    j_{IJ} = \frac{\pi_{I}\cdot p_{I\rightarrow J}}{\langle t_{I}\rangle} - \frac{\pi_{J}\cdot p_{J\rightarrow I}}{\langle t_{J}\rangle}
\label{II:EQ11_JSS}
,\end{equation}
as in the Markov case. The sum of all currents leaving state $I$ vanishes in the steady state $\pi$ of a semi-Markov process,
\begin{equation}
    \sum_{J}j_{IJ} = 0
\label{II:EQ12_JSSSum}
,\end{equation} 
which is known as Kirchhoff's law. The formulation of the steady-state current, \autoref{II:EQ11_JSS}, allows the identification of the mean entropy production in the steady state $\braket{\Delta S_{\text{tot}}}$ as
\begin{align}
    \braket{\Delta S_{\text{tot}}} & = T \sum _{I<J} j_{I J} \ln \frac{\pi_{I}\cdot p_{I\rightarrow J}/\braket{t_{I}}}{\pi_{J}\cdot p_{J\rightarrow I}/\braket{t_{J}}}
\label{II:EQ13_S_SS}
.\end{align}
In equilibrium, the entropy production and all currents vanish. Using $j_{IJ}$ in \autoref{II:EQ11_JSS} leads to
\begin{equation}
    \frac{\pi_{I}\cdot p_{I\rightarrow J}}{\langle t_{I}\rangle} = \frac{\pi_{J}\cdot p_{J\rightarrow I}}{\langle t_{J}\rangle},
    \label{II:EQ14_SM_DB}    
\end{equation}
the generalized detailed balance condition for semi-Markov processes. As shown in Appendix A, the arguments in the proof of direction-time independence can be used to derive the detailed-balance condition in \autoref{II:EQ14_SM_DB}. 

For Markov processes, the transition probability and the first moment are given by $p_{I\rightarrow J} = k_{IJ}/\Gamma_I$ and $\langle t_{I}\rangle = 1/\Gamma_I$, respectively. Thus, \autoref{II:EQ11_JSS}, \autoref{II:EQ13_S_SS} and \autoref{II:EQ14_SM_DB} reduce to the known relations for the Markov case.

\section{Semi-Markov TUR}\label{sec:3}
The DTI condition for semi-Markov kernels, \autoref{II:EQ9_DTI} , allows us to derive a thermodynamic uncertainty relation for semi-Markov processes initialized in the steady state $p_{I_0}(0) = \pi_{I_0}$. This TUR can be formulated as 
\begin{equation}
\mathcal{Q}^I \equiv \frac{\Braket{\mathcal{J}}^2}{T\text{ Var } \mathcal{J}}\frac{2 \tau}{\exp \left( \braket{\Delta S_{\text{tot}}} \tau/T \right) - 1} \leq 1
\label{III:EQ1_tur1} 
.\end{equation} The inverse time scale $1/\tau \equiv \sum_j \pi _j/\braket{t_j}$ measures the frequency of transitions in the steady state.

The proof follows an information inequality approach to TURs \cite{dechant2018, hasegawa2019, dechant2020} centered around the Cramér-Rao inequality
\begin{equation}
\frac{\left( \left. \partial _{\theta}\Braket{f}^\theta \right|_{\theta = 0}\right)^2}{\text{Var } f} \leq \mathcal{I}
\label{III:EQ3_cr} 
,\end{equation} which is valid for general random variables $f$ defined on a family of probability densities $\mathcal{P}^\theta$. The expectation value with respect to $\mathcal{P}^\theta$ is denoted by $\Braket{\cdot}^\theta$. The Fisher information $\mathcal{I}$, defined as
\begin{equation}
    \mathcal{I} \equiv \lim_{\theta \to 0} \frac{2}{\theta^2} \Braket{\ln \frac{\mathcal{P}^\theta\left[\Gamma\right]}{\mathcal{P}\left[\Gamma\right]}}^\theta
\label{III:EQ3_fisI}  
,\end{equation}
measures the infinitesimal Kullback-Leibler distance of a perturbed path weight $\mathcal{P}^\theta \left[\Gamma\right]$ to the original path weight 
$\mathcal{P} \left[\Gamma\right]$ at $\theta = 0$. The perturbed path weight is not connected to a real physical process but rather an auxiliary tool to deduce \autoref{III:EQ1_tur1} from \autoref{III:EQ3_cr}. It is tailor-made to scale stochastic currents as
\begin{equation}
\left. \partial _{\theta}\Braket{\mathcal{J}}^{\theta} \right|_{\theta = 0} = \Braket{\mathcal{J}}
\label{III:EQ4_scaling} 
,\end{equation} which can be substituted into \autoref{III:EQ3_cr} by choosing $f = T \mathcal{J}$. The resulting uncertainty bound can be written in terms of the entropy production by bounding it against the Fisher information $\mathcal{I}$ in the form 
\begin{equation}
\frac{\Braket{\mathcal{J}}^2}{T \text{ Var } \mathcal{J}} \leq
    \frac{1}{T} \mathcal{I} \leq \frac{1}{2 \tau} \left( \exp\left(\braket{\Delta S_{\text{tot}}} \tau/T\right) - 1 \right)
\label{III:EQ5_ineq} 
.\end{equation}
The full proof of \autoref{III:EQ1_tur1}, which includes the details of the perturbed path weight $\mathcal{P}^\theta \left[\Gamma\right]$, is given in Appendix B. The idea of the proof is contained in the concept of the embedded Markov chain. Every semi-Markov process contains a discrete-time Markov chain by forgetting about the time spent in each state. By virtue of the defining \autoref{II:EQ10_S} for semi-Markov processes with DTI, the entropy productions of the semi-Markov process and its corresponding discrete-time Markov chain coincide, as the time spent in a state is symmetric under time reversal. 

By designing a virtual perturbation $\mathcal{P}^\theta \left[\Gamma\right]$ based on the discrete-time result by Proesmans and van den Broeck \cite{proesmans2017}, the modifications to the discrete-time path weight of the embedded Markov chain can be transferred to the semi-Markov path weight without affecting the time-symmetric waiting time distributions $\psi_I(t)$. This virtual perturbation can then be used in \autoref{III:EQ3_cr}.


For a singular waiting time distribution of the form $\psi_I(t) = \delta (t - \tau)$, the derived bound \eqref{III:EQ1_tur1} becomes the discrete-time result for a Markov chain from \cite{proesmans2017}. Our result complements an extant TUR for semi-Markov processes \cite{vu2020}, which relies on a different virtual perturbation $\mathcal{P}^\theta \left[\Gamma\right]$. Compared to the result in \cite{vu2020}, \autoref{III:EQ1_tur1} is exponential rather than linear in the entropy production rate and remains valid for arbitrary waiting time distributions without becoming singular for a delta-like waiting time distribution. Moreover and in contrast to the memory term in the result derived in \cite{vu2020}, all quantities in \autoref{III:EQ1_tur1} can, in principle, be evaluated directly by evaluating currents in systems with a semi-Markov description. Thus, the TUR in \autoref{III:EQ1_tur1} is accessible from an operational point of view.


\section{Markov-based TUR}\label{sec:4}

\begin{figure}[b]
\centering
\includegraphics[width=\linewidth]{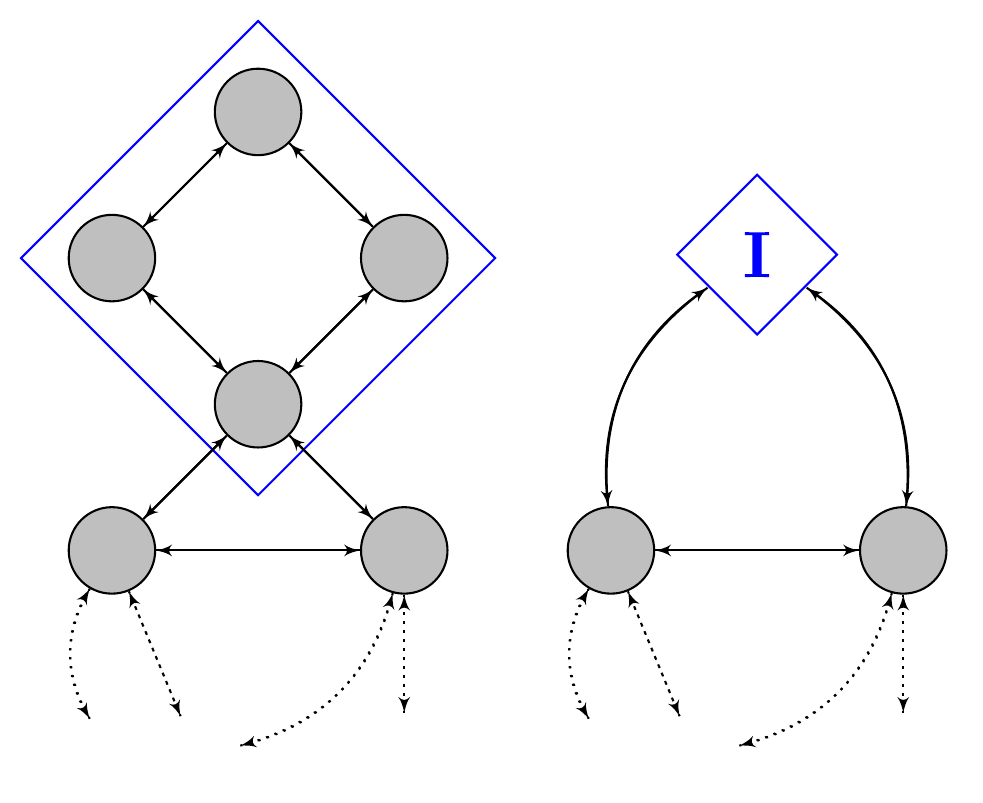}
\caption[cg1]{Section of a Markov network consisting of states (circles) and links (arrows). The diamond indicates a subset in which a single state is connected to the remaining network. In the coarse-grained network the four states in the diamond are treated as a single compound state $I$ with semi-Markovian dynamics.}
\label{IV:FigTheo1_CG1} 
\end{figure}

\begin{figure}[b]
\includegraphics[width=\linewidth]{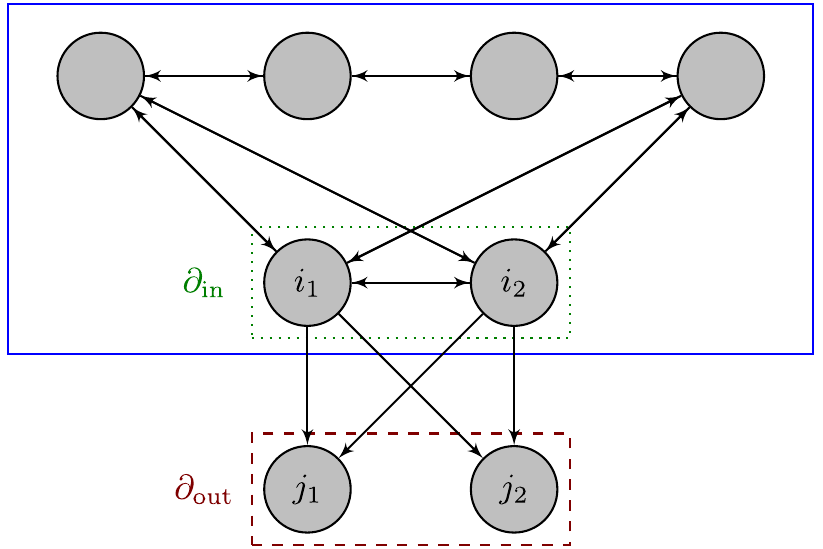}
\caption[cg2]{A coarse-grained compound state $I$ (box). Its outer and inner boundary, $\partial_{\text{out}} I = \{j_1, j_2 \}$ and $\partial_{\text{in}}I = \{i_1, i_2 \}$ are highlighted by a dashed and dotted box, respectively. $I$ is semi-Markov if the waiting time distribution $\widetilde{\psi}_{i \to j}(t)$ of reaching $j = j_1, j_2$ after time $t$ is the same for the starting points $i = i_1$ and $i = i_2$.}
\label{IV:FigTheo2_CG2} 
\end{figure}

A semi-Markov process can originate from coarse-graining an underlying Markov process. However, typically, coarse-graining such a process does not lead to a description satisfying the semi-Markov property. In this section, we consider coarse-graining which lumps multiple Markov states together into a compound semi-Markov state, which satisfies the DTI condition by construction. This type of coarse-graining has been used in \cite{Qian2007} and \cite{martinez2019}, for example.

\subsection{Notation and set-up}

We distinguish semi-Markov states and the more fundamental Markov states by using uppercase and lowercase letters, respectively. Moreover, we consider a connected network and assume that for all Markov states $i, j$, $k_{ij} > 0$ implies $k_{ji} > 0$ , which is necessary to ensure that the backward trajectory is well defined. A group of states $I$ can be regarded as a semi-Markov state if the semi-Markov kernel defined in \autoref{II:EQ1_Psi}, $\psi_{I\to J}(t)$, is well defined, i.e., the history before the system enters $I$ does not affect the joint distribution of its waiting time $t$ and its next destination $J$. 

For example, a group of states $I$ in which only a single state $i \in I$ is connected to states in the remaining network can be treated as a single semi-Markov state because it satisfies the condition of well-definedness readily. This situation is depicted in \autoref{IV:FigTheo1_CG1}, which also illustrates the possibility of cycles hidden within a semi-Markov state. 

Even if several states in a group are connected to states outside, this group can still possibly be described as a semi-Markov state. Before formulating the condition ensuring the well-definedness of the corresponding semi-Markov kernel $\psi_{I\to J}(t)$, some notation is introduced. Let $\partial_{\text{in}}I$ denote the set of boundary states of $I$ that have a link to a state $j \notin I$. States in $\partial_{\text{in}}I$ are the possible states of the system immediately after the coarse-grained description registered a jump into $I$. Similarly, the set of states outside $I$ which can be reached from a state $i \in I$ is denoted $\partial_{\text{out}} I$. The network in \autoref{IV:FigTheo2_CG2} illustrates these definitions. 

The kernel $\psi_{I\to J}(t)$ encodes the distribution of the waiting time in the semi-Markov state $I$ until the system reaches the next state $J$. On the Markov level, the waiting time in $I$ is given by the time spent between visiting a state $i \in \partial_{\text{in}}I$ and reaching an outside state $j \in \partial_{\text{out}} I$. The associated waiting time distribution, denoted $\widetilde{\psi}_{i \to j}(t)$, is obtained by solving an appropriate escape problem \cite{Qian2007}. We consider the network consisting of all states and links within $I$ to which $\partial_{\text{out}} I$ and the unidirectional links from states in $I$ to states in $\partial_{\text{out}} I$ are added. An example is depicted in \autoref{IV:FigTheo2_CG2}. The waiting time distributions $\widetilde{\psi}_{i \to j}(t)$ are calculated as
\begin{equation}
    \widetilde{\psi}_{i \to j}(t) = \frac{\text{d}}{\text{d}t} p_j(t)
\end{equation}
after solving the associated absorbing master equation inside $I$ for the initial condition $p_k(0) = \delta_{ki}$.

The  waiting time distribution $\widetilde{\psi}_{i \to j}(t)$ is precisely the kernel $\psi_{I\to J}(t)$. Thus, the semi-Markov property and also the DTI condition are ensured if $\widetilde{\psi}_{i \to j}(t)$ does not depend on the starting point $i \in \partial_{\text{in}}I$ \cite{Qian2007}. This condition ensures that the previous state from which $I$ is entered does not influence the semi-Markov kernel $\psi_{I \to J}(t)$.

\subsection{Markov-based TUR}
Solely from the existence of an underlying Markovian description, deductions can be made about the nature of the coarse-grained semi-Markov process. This proposition will be discussed from the perspective of TURs.

Markov networks satisfy the ordinary TUR for stochastic currents $\mathcal{J}$ in the form
\begin{equation}
    \mathcal{Q}^{0} \equiv \frac{\braket{\mathcal{J}}^2}{\text{Var }\mathcal{J}} \frac{2}{\braket{\Delta S_{\text{tot}}^{0}}} \leq 1
\label{IV:EQ0_OTUR}
\end{equation}
in the steady state \cite{barato2015, gingrich2016}, where $\langle\Delta S_{\text{tot}}^{0}\rangle$ is the entropy production of the Markov network. This TUR holds for the underlying Markov network in any coarse-grained process. 

For stochastic currents $\mathcal{J}$ that are well-defined within the coarse-grained semi-Markov process, we will prove the Markov-based TUR 
\begin{equation}
\mathcal{Q}^{II} \equiv \frac{\Braket{\mathcal{J}}^2}{\text{Var } \mathcal{J}} \frac{2 }{\Braket{\Delta S_{\text{tot}}}} \leq 1,
\label{IV:EQ1_quality} 
\end{equation} with the mean semi-Markov entropy production $\Braket{\Delta S_{\text{tot}}}$ as defined in \autoref{II:EQ13_S_SS}. The result is valid in the non-equilibrium steady state (NESS) for semi-Markov processes obtained by coarse-graining as described above. The proof is given in Appendix C.

This bound is stronger than \autoref{IV:EQ0_OTUR} since with
\begin{equation}
    \Braket{\Delta S_{\text{tot}}} \leq \Braket{\Delta S_{\text{tot}}^{0}}
\end{equation}
the full entropy production of a Markov network is bounded from below by the entropy production of the coarse-grained description \cite{seifert2019}. This bound is also stronger than the semi-Markov TUR \eqref{III:EQ1_tur1}, as
\begin{equation}
    \mathcal{Q}^I \leq \mathcal{Q}^{II}
\end{equation}
follows from \autoref{III:EQ1_tur1} and the inequality $\exp(x) - 1 \geq x$.

The bound \eqref{IV:EQ1_quality} is only valid for currents $\mathcal{J}$ that are well defined within the semi-Markov dynamics. In the example of \autoref{IV:FigTheo1_CG1}, jumps along the links on the lower three-state cycle can be included in $\mathcal{J}$, whereas jumps within $I$ cannot. In particular, contributions from hidden cycles as in \autoref{IV:FigTheo1_CG1} contribute to $\braket{\Delta S_{\text{tot}}^{0}}$ but not to $\braket{\Delta S_{\text{tot}}}$. 

\begin{figure*}[bt]
    \begin{subfigure}[t]{0.32\textwidth}
      \centering
        \includegraphics[width=\linewidth]{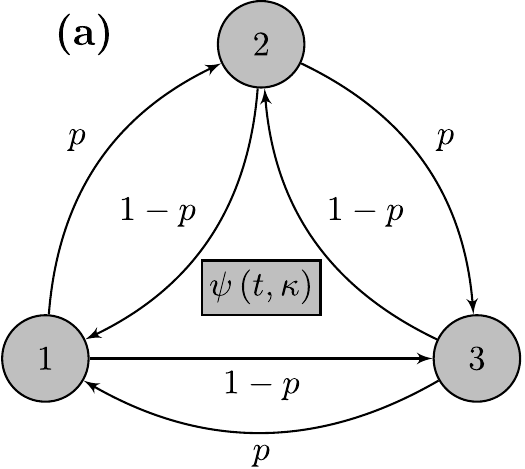}
    \end{subfigure}
    \begin{subfigure}[t]{0.32\textwidth}
      \centering
        \includegraphics[width=\linewidth]{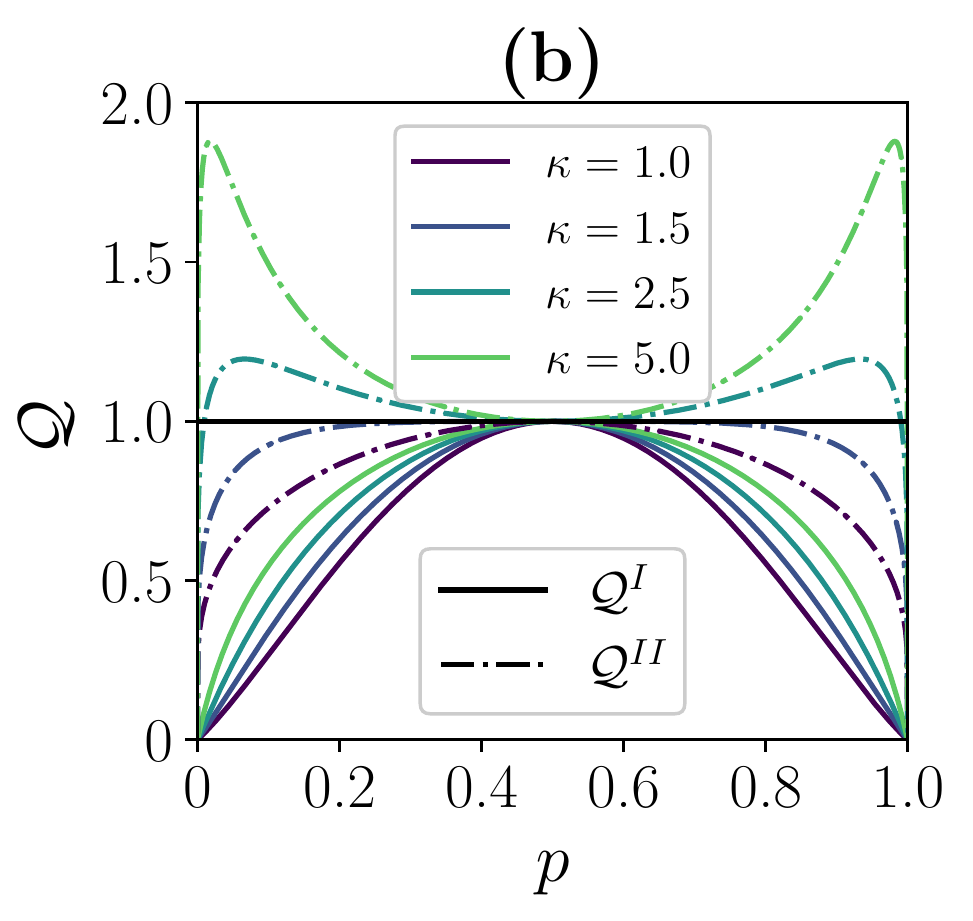}
    \end{subfigure}    
    \hfill 
    \begin{subfigure}[t]{0.32\textwidth}
      \centering
        \includegraphics[width=\linewidth]{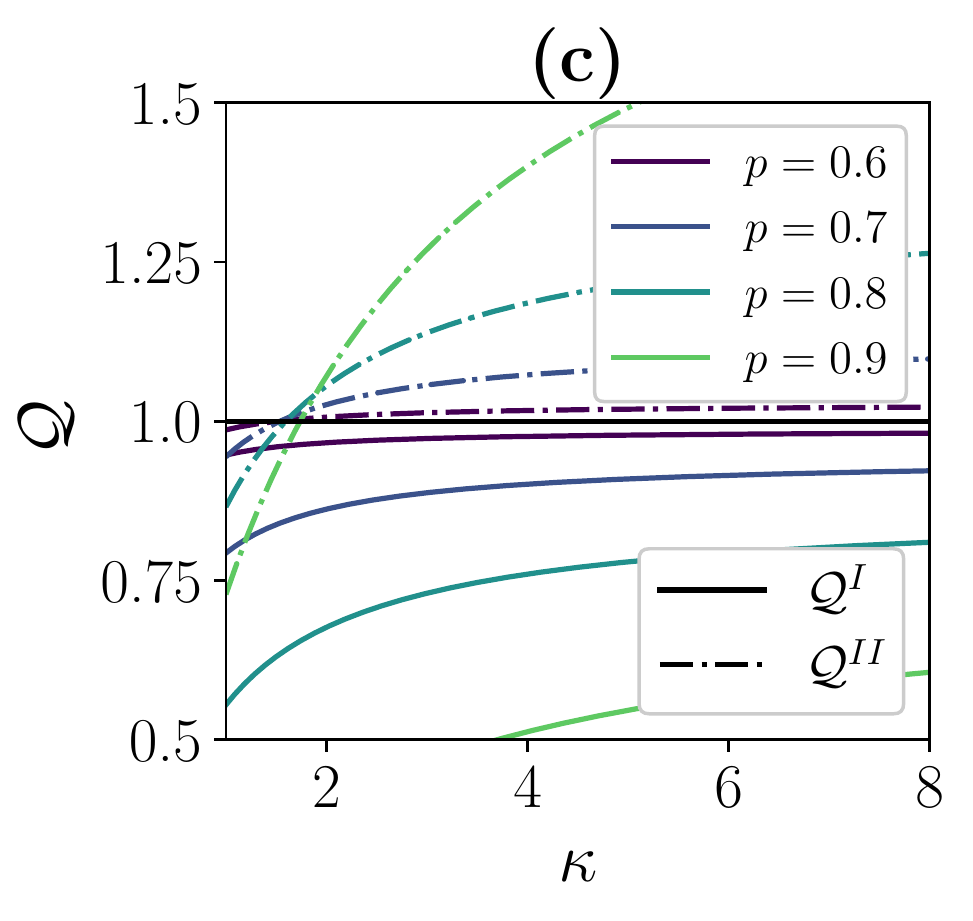}
    \end{subfigure} 
    \caption[TUR Random-walk]{(a): Asymmetric random-walk on three states with identical waiting-time distributions $\psi\left(t,\kappa\right)$, \autoref{IV_Num:EQ1_Gamma}, and transition probabilities $p$ and $1-p$. (b) and (c): TUR quality factors $\mathcal{Q}^{I}$ and $\mathcal{Q}^{II}$ as a function of $p$ for fixed $\kappa$ and $\kappa$ for fixed $p$, respectively, increasing from dark color to bright color.}
\label{V:Fig1_Network} 
\label{IV:Fig2_Scatter}
\end{figure*}

A further consequence of the bound \eqref{IV:EQ1_quality}, when contrasted to the semi-Markov TUR in \autoref{III:EQ1_tur1}, is that not all semi-Markov processes can be the result of coarse-graining a Markov process. As an illustration, consider the concrete example of a singular kernel $\psi_{I \to J}(t) = \delta(t - \Delta T) p_{I \to J}$. This waiting time distribution imitates a discrete time step $\Delta T$. It is known that a TUR of the form \eqref{IV:EQ0_OTUR} does not hold for discrete-time Markov processes in general \cite{shiraishi2017, dechant2020}. Consequently, these processes violate the bound \eqref{IV:EQ1_quality} and thus cannot arise from coarse-graining a continuous-time Markov network. This holds true notwithstanding the fact that mathematically, i.e. without imposing thermodynamic consistency, it is known how to approximate singular waiting time distributions with Markov states \cite{david1987}. 

\subsection{Thermodynamic inference scheme for identifying underlying Markov networks}\label{Sec:Inf}
The range of validity of the Markov-based TUR \eqref{IV:EQ1_quality} can be used to infer whether a semi-Markov process possesses an underlying Markov network or not. As discussed above, the Markov-based TUR is established for semi-Markov processes emerging from coarse-grained Markov networks. If the Markov-based TUR is violated for a semi-Markov process, this process cannot have an underlying Markov network. Operationally, the following inference scheme can be used to identify underlying Markov networks. 
The direction-time independence of the process has to be verified in the first step. For this purpose, all semi-Markov kernels $\psi_{I\rightarrow J}\left(t\right)$ for all states $I$ are needed. If the ratio 
\begin{equation}
    \frac{\psi_{I\rightarrow J}\left(t\right)}{\psi_{I\rightarrow K}\left(t\right)}
\label{IV_Inf:EQ1_Ratio}
\end{equation}
is independent of $t$ for all possible destinations $K$ and all $J$, the process is direction-time independent. If this condition is not satisfied, the process cannot emerge from coarse graining as presented above \cite{Qian2007}.

The quality factor $\mathcal{Q}^{II}$ of the Markov-based TUR can be calculated for semi-Markov processes with direction-time independence. A violation of this TUR implies that the investigated semi-Markov process cannot result from coarse graining an underlying Markov network, although it may be thermodynamically consistent.

It is important to note that the converse of the described criterion does not hold. If $\mathcal{Q}^{II} \leq 1$, the investigated process does not need to have an underlying Markov network.

\section{Illustration and numerical evidence}\label{sec:5}

We illustrate our results with two concrete examples. For both TURs, we consider the quality factors $\mathcal{Q}^{I}$ and $\mathcal{Q}^{II}$ as defined by the semi-Markov TUR, \autoref{III:EQ1_tur1}, and the Markov-based TUR, \autoref{IV:EQ1_quality}, respectively.

\subsection{Asymmetric random walk}

First, we consider an asymmetric random walk between three semi-Markov states as shown in \autoref{V:Fig1_Network} (a). We assume that the waiting time distribution for all states is a gamma distribution of the form
\begin{equation}
    \psi\left(t, \kappa\right) = \kappa \frac{(\kappa t)^{\kappa-1}e^{-\kappa t}}{\Gamma\left(\kappa \right)},
    \label{IV_Num:EQ1_Gamma}
\end{equation}
with shape parameter $\kappa$, scale parameter $1/\kappa$ and the gamma function $\Gamma\left(\kappa \right)$. For all $\kappa$, the first moment of the distribution is fixed to one, thus the time scale $\tau$ in \autoref{III:EQ1_tur1} is equal to one as well.

The waiting time distribution in \autoref{IV_Num:EQ1_Gamma} interpolates between two different classes of systems. For $\kappa = 1$, the system becomes a Markov random walk. For $\kappa\rightarrow\infty$, the waiting time distribution $\psi\left(t,\kappa\right)$ becomes a delta function, hence the system becomes a discrete-time Markov chain. For values of $\kappa$ in-between, the system is a semi-Markov random walk.

\begin{figure*}[t]
     \centering
     \begin{subfigure}[t]{0.48\textwidth}
        \centering
        \includegraphics[width=\linewidth]{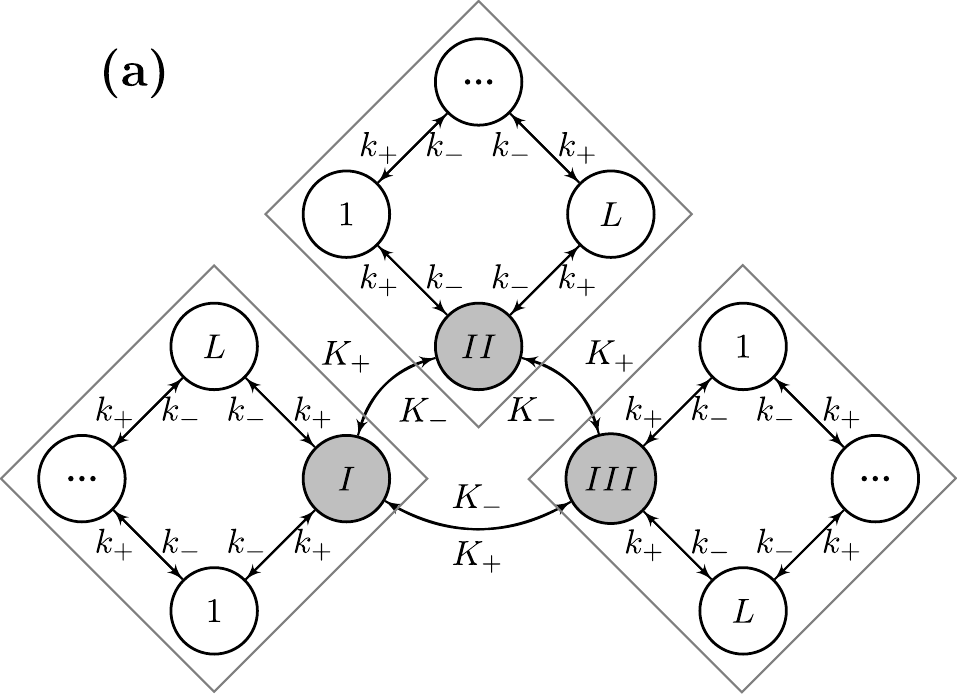}
     \end{subfigure}
     \begin{subfigure}[t]{0.49\textwidth}
         \centering
         \includegraphics[width=\textwidth]{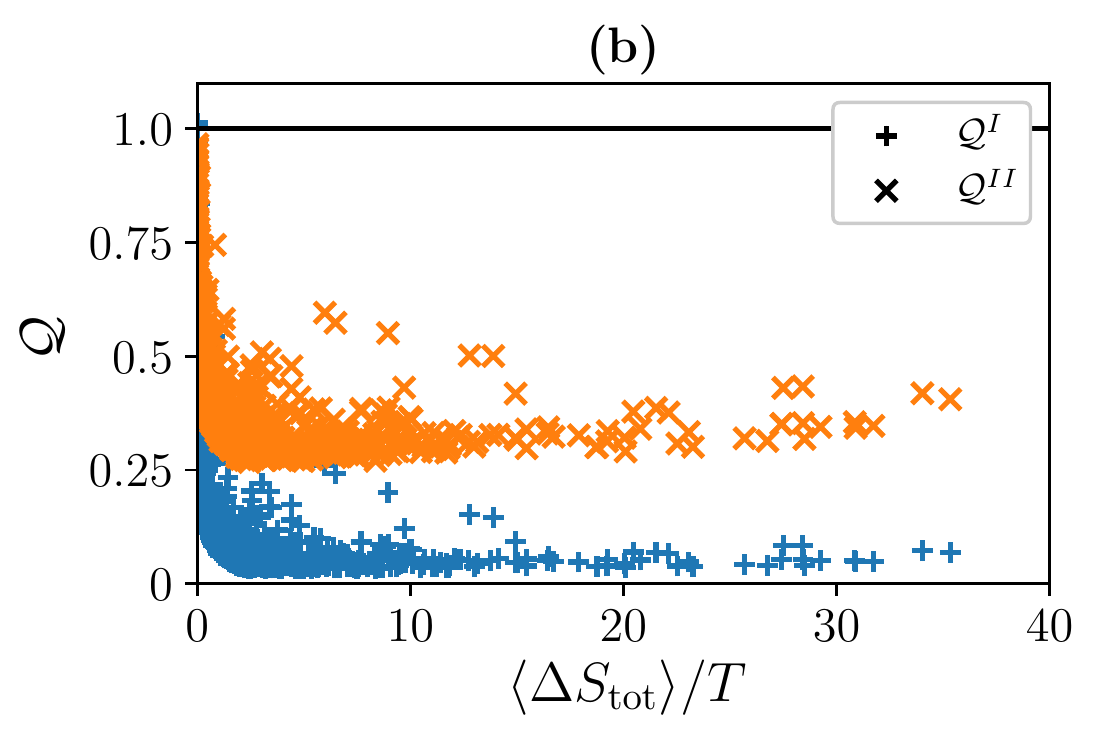}
     \end{subfigure}
     \caption[TUR multi-ring model]{(a): Multi-ring model with three coarse-grained mesostates. Each mesostate contains a sub-ring with length $L$, highlighted with a grey box. The transition rates along the main ring in forward and backward direction are $K_{+}$ and $K_{-}$, respectively. Corresponding rates in the sub-rings are denoted $k_{+}$ and $k_{-}$. (b): Simulated quality factors $\mathcal{Q}^{I}$ and $\mathcal{Q}^{II}$ for both TURs as functions of the mean entropy production rate $\langle\Delta S_{\text{tot}}\rangle/T$ for 800 realizations of the system for $K_{-} = k_{-} = 1$. $L$ is uniformly distributed between 1 and 1000 and the parameter $k = K_{+} = k_{+}$ is drawn randomly between 1 and 35 from a uniform distribution.}
     \label{SV:FIG1_Ring}
\end{figure*}

If we choose the net number of clockwise jumps between the states as the current $\mathcal{J}$, both quality factors will only depend on $p$ and $\kappa$. As calculated in \cite{Maes2009}, the mean and variance of this current are given by
\begin{equation}
    \langle\mathcal{J}\rangle = 2p - 1
    \label{IV_Num:EQ2_JMean}
\end{equation}
and 
\begin{equation}
    T \text{ Var } \mathcal{J} = \left(1/\kappa - 1\right)\left(2p - 1\right)^2 + 1
    \label{IV_Num:EQ3_JVar}
,\end{equation}
respectively. Using \autoref{II:EQ13_S_SS}, the mean entropy production rate can be calculated as
\begin{equation}
    \langle\Delta S_{\text{tot}}\rangle/T = \left(2p-1\right)\ln\frac{p}{1-p}
    \label{IV_Num:EQ4_Ent}
.\end{equation}

The quality factors $\mathcal{Q}^{I}$ and $\mathcal{Q}^{II}$ are shown as a function of $p$ and $\kappa$ in \autoref{V:Fig1_Network} (b) and \autoref{V:Fig1_Network} (c) respectively, with the other parameter kept fixed. Both quality factors approach $1$ near equilibrium, i.e., for $p = 0.5$. While $\mathcal{Q}^I \leq 1$ is always satisfied by virtue of the semi-Markov TUR, \autoref{III:EQ1_tur1}, $\mathcal{Q}^{II}$ becomes larger than 1 for $\kappa > \kappa_c(p)$ with some threshold value $\kappa_c(p)$, which implies a violation of the Markov-based TUR. Therefore, the observed semi-Markov process cannot emerge from an underlying Markov-process by lumping states together. Note that reversing the argument is not true, i.e., the system potentially does not have an underlying Markov network even for $\kappa$ small enough to ensure $\mathcal{Q}^{II}\leq 1$.

\subsection{Multi-ring model}

As a second illustration, we consider the multi-ring network shown in \autoref{SV:FIG1_Ring} (a), which is investigated numerically. This system with three semi-Markov states emerges from coarse-graining an underlying Markov network. All states have identical semi-Markov kernels with DTI included by construction.

The quality factors $\mathcal{Q}^{I}$ and $\mathcal{Q}^{II}$ can be calculated by simulating the underlying Markov network with the Gillespie algorithm \cite{Gillespie_1977}, since the quantities for both TURs are accessible in the statistics. The transition probabilities and the first moment suffice to calculate the mean entropy production. These quantities can be determined from counting jumps along the main ring and averaging the residence times in the sub-rings. The chosen current is the net number of jumps along the main ring, which is obtained by counting. By symmetry, the time constant $\tau$ in the semi-Markov TUR is equal to the first moment of the waiting time distribution.

Both quality factors $\mathcal{Q}^{I}$ and $\mathcal{Q}^{II}$ are shown in \autoref{SV:FIG1_Ring} (b) as a function of the mean entropy production rate $\langle\Delta S_{\text{tot}}\rangle/T$ for randomly drawn $L$ and $k = K_{+} = k_{+}$. In accordance with the TUR bounds, both quality factors are smaller than one. Moreover, the Markov-based TUR, which makes use of the underlying Markov description, is tighter than the semi-Markov TUR.

\section{Conclusion}\label{sec:6}

In this paper, we have discussed the thermodynamically consistent description of semi-Markov processes. We have introduced a more intuitive approach to the past average based on a discrete approximation of the semi-Markov kernel. In addition, an alternative derivation of the direction-time independence criterion from \cite{chari1994} for semi-Markov kernels has been presented. 

As one main result, we have proven a thermodynamic uncertainty relation valid for semi-Markov processes with direction-time independence. This uncertainty relation generalizes the discrete-time result from \cite{proesmans2017} to semi-Markov processes satisfying DTI. As a second main result, we have shown that a naive generalization of the steady state TUR in Markov networks is valid for semi-Markov processes with underlying Markov structure. For a general semi-Markov process, observing a violation of the Markov-based TUR will exclude the possibility of an underlying Markovian description.

\begin{figure}[h]
\includegraphics[width=\linewidth]{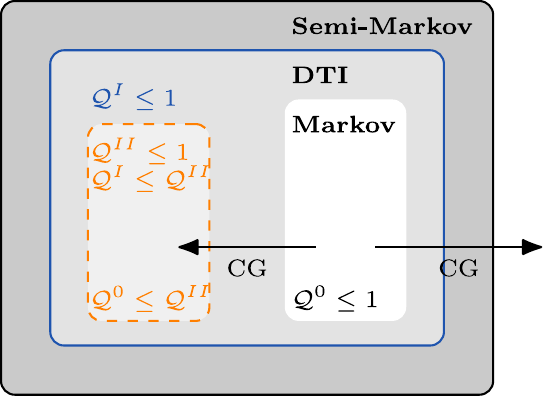}
\caption[Venn]{Venn-type diagram for the hierarchy of the considered stochastic processes and TURs. The blue and orange box illustrate the range of validity of the semi-Markov TUR ($\mathcal{Q}^I \leq 1$) and the Markov-based TUR ($\mathcal{Q}^{II} \leq 1$), respectively. The arrows show different possibilities for coarse-graining: If coarse-graining leads to a semi-Markov process with DTI, the Markov-based TUR is a stronger bound than both the semi-Markov TUR ($\mathcal{Q}^{I} \leq \mathcal{Q}^{II}$)  and the ordinary TUR ($\mathcal{Q}^{0} \leq \mathcal{Q}^{II}$) of the underlying Markov process. Coarse-graining may also leave the realm of semi-Markov processes.}
\label{VI:Fig1_Venn} 
\end{figure}

The hierarchy of the considered processes and inequalities can be summarized in a Venn-type diagram shown in \autoref{VI:Fig1_Venn}. The semi-Markov TUR, \autoref{III:EQ1_tur1}, is valid for the subset of semi-Markov processes fulfilling the DTI criterion. If grouping states together in a Markov network leads to a semi-Markov process, the Markov-based TUR, \autoref{IV:EQ1_quality}, holds true. This bound is tighter than both the semi-Markov TUR and the ordinary TUR, \autoref{IV:EQ0_OTUR}, for the underlying Markov process.

Further work could address the following issues. For Markov processes, the TUR has been generalized to time-dependent driving \cite{koyuk2020}. This generalization could also be possible for the semi-Markov TUR. The formulation of this generalized TUR requires a suitable framework, which has not been introduced yet. Affinity-dependent bounds \cite{pietzonka2016} comprise another possibility for generalizations.

Different coarse-graining strategies not used in this work, especially coarse graining with kinetic hysteresis \cite{hartich2021}, can potentially lead to further TURs with different ranges of validity when paired with appropriately modified techniques. Coarse graining leaves another open question. Semi-Markov processes typically emerge from coarse graining of particularly symmetric systems for which the corresponding escape problem is independent of the starting point to circumvent state memory effects \cite{Qian2007}. Including state memory in a thermodynamically consistent way, based on a semi-Markov model or a different setting, remains an ambitious challenge. 

Lastly, a more detailed analysis or generalization of the proposed inference scheme for underlying Markov networks may yield an understanding of the relation between thermodynamically consistent models of the same system based on different degrees of available information. 

\section*{Acknowledgements}
B.E. and U.S. thank L. Oberreiter for many insightful discussions and integral contributions to the early stages of this work. J.v.d.M. thanks M. Schölpple for helpful technical remarks.

\newpage

\clearpage

\appendix

\widetext

\section{Proof of direction-time independence}
The direction-time independence criterion for thermodynamic consistency of semi-Markov processes in equilibrium was proven in \cite{chari1994}. An alternative, more intuitive proof based on essentially the same argument is presented here. Suppose a generic trajectory $\Gamma$ initialized in the steady state jumps from state $I_{0}$ to state $I_{1}$ at time $t_{1}$ and survives in state $I_{1}$ for at least time $T-t_{1}$. 

The path weight for this trajectory is (cf. equation (2) in the main text)
\begin{equation}
    \mathcal{P}[\Gamma] = \pi_{I_{0}}\frac{\int_{t_{1}}^{\infty}\psi_{I_{0}\rightarrow I_{1}}\left(\tau\right)d\tau}{\langle t_{I_{0}}\rangle}\sum_{J}\int_{T-t_{1}}^{\infty}\psi_{I_{1}\rightarrow J}\left(t\right)dt,
\label{SI:EQ1_OJFor}
\end{equation}
where $\pi_{I_{0}}$ is the equilibrium probability for state $I_{0}$.  The path weight for the backward trajectory $\tilde{\Gamma}$ is given by
\begin{equation}
    \mathcal{P}[\tilde{\Gamma}] = \pi_{I_{1}}\frac{\int_{T-t_{1}}^{\infty}\psi_{I_{1}\rightarrow I_{0}}\left(\tau\right)d\tau}{\langle t_{I_{1}}\rangle}\sum_{J}\int_{t_{1}}^{\infty}\psi_{I_{0}\rightarrow J}\left(t\right)dt,
\label{SI:EQ2_OJBack}
\end{equation}
using equation (9) in the main text for the construction. A thermodynamically consistent process satisfies time-reversal symmetry in equilibrium. If $\Gamma$ is symmetric under time-reversal, the path weights for the forward process $\Gamma$ and the backward process $\tilde{\Gamma}$ are equal. Setting \autoref{SI:EQ1_OJFor} equal to \autoref{SI:EQ2_OJBack} leads to
\begin{equation}
    \pi_{I_{0}}\frac{\int_{t_{1}}^{\infty}\psi_{I_{0}\rightarrow I_{1}}\left(\tau\right)d\tau}{\langle t_{I_{0}}\rangle}\sum_{J}\int_{T-t_{1}}^{\infty}\psi_{I_{1}\rightarrow J}\left(t\right)dt = \pi_{I_{1}}\frac{\int_{T-t_{1}}^{\infty}\psi_{I_{1}\rightarrow I_{0}}\left(\tau\right)d\tau}{\langle t_{I_{1}}\rangle}\sum_{J}\int_{t_{1}}^{\infty}\psi_{I_{0}\rightarrow J}\left(t\right)dt
\label{SI:EQ3_Equality}  
,\end{equation}
which should be valid for all times $t_1$ and $T$. 
As a consequence, the integrals should be equal for any time domain expressed in terms of $t_{1}$ and $T$. Therefore, the integrals on the left and on the right side of \autoref{SI:EQ3_Equality} can be replaced by their integrands evaluated at their lower boundary. This results in
\begin{equation}
    \frac{\pi_{I_{0}}}{\langle t_{I_{0}}\rangle}\psi_{I_{0}\rightarrow I_{1}}\left(t_{1}\right)\psi_{I_{1}}\left(T-t_{1}\right) = \frac{\pi_{I_{1}}}{\langle t_{I_{1}}\rangle}\psi_{I_{1}\rightarrow I_{0}}\left(T-t_{1}\right)\psi_{I_{0}}\left(t_{1}\right)
    \label{SI:EQ4_Abl1}     
,\end{equation}
with the waiting time distribution $\psi_{I}\left(t\right) = \sum_{J}\psi_{I\rightarrow J}\left(t\right)$ defined as in the main text. As \autoref{SI:EQ4_Abl1} is satisfied for all $t_1$ and $T$, the ratio of semi-Markov kernel and waiting time for the states $I_{0}$ and $I_{1}$ has to be independent of both $t_1$ and $T$. Since the trajectory $\Gamma$ is generic, this criterion holds for any semi-Markov state $I$. 
From
\begin{equation}
    \frac{\psi_{I\rightarrow J}\left(t\right)}{\psi_{I}\left(t\right)} = \frac{\psi_{I\rightarrow J}\left(t\right)}{\sum_{J}\psi_{I\rightarrow J}\left(t\right)} \overset{!}{=} \alpha_{I\to J}
\label{SI:EQ5_Abl2}   
\end{equation}
with some function $\alpha_{I\to J}$ independent of $t$, the separability of the semi-Markov kernel
\begin{equation}
    \psi_{I\rightarrow J}\left(t\right) = \alpha_{I\rightarrow J}\psi_{I}\left(t\right)
\label{SI:EQ6_Abl3}     
\end{equation}
can be deduced. By imposing normalization conditions for semi-Markov kernel and waiting time distribution as in the main text, $\alpha_{I\rightarrow J}$ can be identified as the transition probability $p_{I\rightarrow J}$. This leads to
\begin{equation}
    \psi_{I\rightarrow J}\left(t\right) = p_{I\rightarrow J}\psi_{I}\left(t\right),
\label{SI:EQ7_DTI}     
\end{equation}
the direction-time independence criterion for semi-Markov process in equilibrium as established in \cite{chari1994}. As shown there, the generalized detailed balance condition for semi-Markov processes is also included in this argument. Since \autoref{SI:EQ3_Equality} is valid for all times, specifying $t_{1} = T - t_{1} = 0$ in \autoref{SI:EQ3_Equality} results in
\begin{equation}
    \frac{\pi_{I_{0}}p_{I_{0}\rightarrow I_{1}}}{\langle t_{I_{0}}\rangle} = \frac{\pi_{I_{1}}p_{I_{1}\rightarrow I_{0}}}{\langle t_{I_{1}}\rangle},
    \label{SI:EQ8_DTI} 
\end{equation}
the generalized detailed balance condition for semi-Markov processes as in \cite{Qian2007,Maes2009}. This is not surprising given that the path weight for a one-jump trajectory without waiting times in fact describes a jump of a semi-Markov process. 


\section{Proof of the semi-Markov TUR}
In this appendix, we prove the uncertainty relation referred to as semi-Markov TUR in the main text.

\subsection{Preliminaries}
The proof follows the idea of a virtual perturbation to the path weight $\mathcal{P}\left[\Gamma\right] \to \mathcal{P}^\theta\left[\Gamma\right]$, where the modified path weight $\mathcal{P}^\theta\left[\Gamma\right]$ takes the role of a comparison dynamics to the original process. The notation follows the main text, i.e., the original path weight (at $\theta = 0$) is given by 
\begin{eqnarray}
    \mathcal{P}\left[\Gamma\right] &&= p_{I_{0}}\left(0\right)\Psi_{I_{0}\rightarrow I_{1}} \left(t_{1}\right) \prod_{i=1}^{L-1} \left(\psi_{I_{i}}\left(t_{i+1}-t_{i}\right) p_{I_{i}\rightarrow I_{i+1}}\right) \Phi_{I_{L}}\left(T-t_{L}\right)
\label{SM:tur1:1}
,\end{eqnarray}
following the definition of the semi-Markov path weight by imposing the DTI condition, 
\autoref{SI:EQ7_DTI}. As argued in the main text, we allow the jump probabilities to depend on $\theta$, which is written as $p_{I_{i} \to I_{i+1}}^\theta$. The waiting time distributions $\psi_{I_{i}}\left(t_{i+1}-t_{i}\right)$ and the steady state $p_{I_0} = \pi_{I_0}$ remain unchanged. Note that the past average term $\Psi_{I_{0}\rightarrow I_{1}} \left(t_{1}\right)$ becomes $\theta$-dependent as well, as it depends on $p_{I_{0} \to I_{1}}$ via $\psi_{I_{0}\rightarrow I_{1}}(t)$. The difference in the path weights can be highlighted by using \eqref{SM:tur1:1} for the log-ratio of $\mathcal{P}^\theta\left[\Gamma\right]$ to $\mathcal{P}\left[\Gamma\right]$,
\begin{equation}
    \ln \frac{\mathcal{P}^\theta\left[\Gamma\right]}{\mathcal{P}\left[\Gamma\right]} = \sum_{i = 0}^{L-1} \ln \frac{p^\theta_{I_{i} \to I_{i+1}}}{p_{I_{i} \to I_{i+1}}} = \sum_{I J} n_{IJ} \ln  \frac{p^\theta_{I \to J}}{p_{I \to J}}
\label{SM:tur1:2}  
,\end{equation} where the $i=0$ term stems from the past average term $\Psi_{I_{0}\rightarrow I_{1}} \left(t_{1}\right)$. As in the main text, $n_{IJ}$ is the number of jumps in the direction $I \to J$ along the trajectory. 

\subsection{Conditions on the comparison dynamics}\label{sec:conditions}
The comparison dynamics is supposed to satisfy the following three properties. These design principles impose conditions on the jump probabilities $p_{I \to J}^\theta$ of the modified process.

\begin{enumerate}
    \item Semi-Markov currents $\mathcal{J}$ are scaled in the form $\left. \partial _\theta \right|_{\theta = 0} \braket{\mathcal{J}}^\theta = \braket{\mathcal{J}}^0$. This condition is ensured by scaling the NESS probability current between two states $I$ and $J$
    \begin{equation}
        j_{IJ}^\theta = \frac{1}{T} \left( \braket{n_{IJ}}^{\theta} - \braket{n_{JI}}^{\theta} \right) = \frac{\pi_I}{\braket{t_I}}p_{I \to J}^\theta  - \frac{\pi_J}{\braket{t_J}} p_{J \to I}^\theta
        \label{SM:tur1:4}  
    \end{equation}
    according to
    \begin{equation}
        \partial _\theta j_{IJ}^\theta = j_{IJ}
        \label{SM:tur1:3}  
    .\end{equation} For Markov processes, $p_{I \to J} = k_{IJ}/\Gamma_I$ and $\braket{t_I} = 1/\Gamma_I$ restore the known formula for the steady state probability current from \autoref{SM:tur1:4}.  
    \item The comparison dynamics respects the stationary state, i.e., $\pi ^\theta_I = \pi _I$. 
    This condition is met because scaling all currents by a factor is consistent with keeping the stationary state: From the defining property of the stationary state,
    $0 = -\partial_t \pi_I^\theta = \sum _J j_{IJ} ^\theta$, we can expand to first order in $\theta$ using \autoref{SM:tur1:3}. The resulting equations, $0 = -\partial_t \pi_I^\theta = (1 + \theta) \sum_J j_{IJ}$ are satisfied for $\pi_I^\theta = \pi_I$ because $0 = -\partial _t \pi_I = \sum_J j_{IJ}$.
    \item The comparison dynamics does not change the waiting times, i.e., $\braket{t_I}^\theta = \braket{t_I}$.
    This condition is satisfied since the waiting time distribution $\psi_I(t)$ is not affected by $\theta$.
\end{enumerate}

\subsection{The proof}
By substituting $f = T \mathcal{J}$ into the Cramér-Rao inequality,
\begin{equation}
\frac{\left( \left. \partial _{\theta}\Braket{\mathcal{J}}^{\theta} \right|_{\theta = 0}\right)^2}{\text{Var } \mathcal{J}} \leq \frac{\left( \left. \partial _{\theta}\Braket{f}^{\theta} \right|_{\theta = 0}\right)^2}{\text{Var } f} \leq \mathcal I
    \label{SM:tur1:cr0}
,\end{equation}
we see that the tightest TUR bound is obtained for the smallest Fisher information $\mathcal{I}$ that is consistent with the conditions of the previous section. We can calculate the Fisher information from its definition
\begin{equation}
    \mathcal{I} \equiv \lim_{\theta \to 0} \frac{2}{\theta^2} D_{KL}(\theta||0) = \lim_{\theta \to 0} \frac{2}{\theta^2} \Braket{\ln \frac{\mathcal{P}^\theta\left[\Gamma\right]}{\mathcal{P}\left[\Gamma\right]}}^\theta = \lim_{\theta \to 0} \frac{2}{\theta^2} \sum_{IJ} \braket{n_{IJ}}^\theta \ln \frac{p^\theta_{I \to J}}{p_{I \to J}} 
    \label{SM:tur1:fisher}
\end{equation}
by using \autoref{SM:tur1:2} for the last equality. In the steady state, the expected number of jumps along $I \to J$ (not counting the reverse) is
\begin{equation}
    \braket{n_{IJ}}^\theta = \frac{\pi_I}{\braket{t_I}^{\theta}} p_{I \to J}^\theta T = \frac{\pi_I}{\braket{t_I}} T \left( p_{I \to J} + \theta \left. \partial _\theta \right| _{0} p_{I \to J}^\theta + \mathcal{O}(\theta^2) \right)
\label{SM:tur1:jumpn}  
\end{equation} for trajectories of length $T$, since $\braket{t_I}^\theta = \braket{t_I}$ by construction. After substituting \autoref{SM:tur1:jumpn} into \autoref{SM:tur1:fisher}, we  can expand the logarithm in orders of $\theta$,
\begin{align}
    \mathcal{I} & = \lim_{\theta \to 0} \frac{2}{\theta^2} \sum_{IJ} \frac{\pi_I}{\braket{t_I}} T \left( p_{I \to J} + \theta \left. \partial _\theta \right| _{0} p_{I \to J}^\theta + \mathcal{O}(\theta^2) \right) \left[ \theta \frac{\left. \partial _\theta \right| _{0} p_{I \to J}^\theta}{p_{I \to J}} + \theta^2 \left( \frac{\left. \partial^2 _\theta \right| _{0} p_{I \to J}^\theta}{p_{I \to J}} - \frac{1}{2} \left( \frac{\left. \partial _\theta \right| _{0} p_{I \to J}^\theta}{p_{I \to J}} \right)^2 \right) + \mathcal{O}(\theta^3) \right] \nonumber \\
    & = T \sum_{IJ} \frac{\pi_I}{\braket{t_I}} \left( \frac{\left. \partial _\theta \right| _{0} p_{I \to J}^\theta}{p_{I \to J}} \right)^2 p_{I \to J} 
\label{SM:tur1:fisherExplicit}  
.\end{align} Note that, as typical for calculating the Fisher information as a limit from the Kullback-Leibler divergence, normalization arguments have been used. Here, $\partial _\theta \sum _J p_{I \to J}^\theta = \partial_\theta 1 = 0$ implies that contributions proportional to $\theta$ or terms involving the second derivative in the form $\left. \partial^2 _\theta \right| _{0} p_{I \to J}$ vanish.

We now perform a change in notation which effectively shifts attention from the semi-Markov process to its embedded Markov chain \cite{Qian2007}. By introducing the inverse time scale $1/{\tau} \equiv \sum _J \pi_J/\braket{t_J}$,  we can abbreviate
\begin{align*}
p_I \equiv \frac{\pi_I}{\braket{t_I}} \tau, \quad
p_{IJ} \equiv p_I p_{I \to J}, \quad
p_{IJ}^\theta \equiv p_I p_{I \to J}^\theta \quad \text{and} \quad
j_{IJ}^\theta = \frac{1}{\tau} \left( p_{IJ}^\theta  - p_{JI}^\theta \right)
.\end{align*}
The last equation follows from equation \eqref{SM:tur1:4}. The new notation keeps the algebra more transparent by stressing the similarity to the discrete-time result. We see from \autoref{SM:tur1:fisherExplicit} that only the linear dependence on $\theta$ is relevant, thus a simple linear ansatz of the form
\begin{equation}
    p^\theta_{IJ} = p_{IJ} + \theta \left( p_{IJ} - q_{IJ} \right)
    \label{SM:tur1:ansatz1}  
\end{equation}
with $\theta$-independent parameters $q_{IJ}$ is sufficiently general. As mentioned in the main text, the comparison dynamics operates on the embedded discrete-time Markov chain, which can be characterized by the probabilities $p_{IJ}$. Thus, we can use the comparison dynamics by Proesmans and van den Broeck \cite{proesmans2017}, which transferred into this formalism reads as
\begin{align}
    \label{SM:tur1:proesmans} 
    q_{IJ} = & \; \frac{1}{\mathcal{N}} \frac{p_{IJ} p_{JI}}{p_{IJ} + p_{JI}}, \\
    \mathcal{N} := & \;\sum_{IJ} \frac{p_{IJ} p_{JI}}{p_{IJ} + p_{JI}} \nonumber
.\end{align} 
A simple motivation for the ansatz \eqref{SM:tur1:proesmans} is given after completing the proof. Note that the symmetry $q_{IJ} = q_{JI}$ ensures the correct scaling of the probability currents according to \eqref{SM:tur1:3}. In particular, both the stationary state of the embedded discrete-time process, $p_I$, and the stationary state $\pi _I$ of the semi-Markov process remain stationary. Thus, conditions 1 and 2 of Appendix \ref{sec:conditions} are satisfied, while condition 3 is met by construction. 

To conclude the calculation, we rewrite the Fisher information \eqref{SM:tur1:fisherExplicit} as
\begin{equation}
\mathcal{I} = T \sum_{IJ} \frac{\pi_I}{\braket{t_I}} \left( 1 - \frac{q_{IJ}}{p_{IJ}} \right)^2 \frac{p_{IJ}}{p_I} = \frac{T}{\tau} \sum_{IJ}  \left( 1 - \frac{1}{\mathcal{N}} \frac{p_{JI}}{p_{IJ} + p_{JI}} \right)^2 p_{IJ} \\
\label{SM:tur1:fisherAnsatz}  
\end{equation}
in the newly introduced notation. In the first equality, we use the general ansatz \eqref{SM:tur1:ansatz1}, whereas we specify to \eqref{SM:tur1:proesmans} in the second equality. We are then able to confirm
\begin{equation}
\mathcal{I} = \frac{T}{\tau} \left( \frac{1}{2 \mathcal{N}} - 1 \right)
\label{SM:tur1:proesmansI} 
\end{equation}
by direct calculation, which allows us to deduce
\begin{equation}
\Braket{\mathcal{J}} ^2 = \left( \left. \partial _\theta \right| _{0} \Braket{\mathcal{J}}^{\theta} \right)^2 \leq \frac{1}{\tau} \left( \frac{1}{2 \mathcal{N}} - 1 \right) \text{Var } \mathcal{J}
\label{SM:tur1:cr} 
\end{equation}  
from the Cramér-Rao inequality \eqref{SM:tur1:cr0}. The proposed TUR follows by comparing the Fisher information \eqref{SM:tur1:proesmansI} to the steady state semi-Markov entropy production
\begin{equation}
\braket{\Delta S_{\text{tot}}} = T \sum _{I<J} \frac{p_{IJ} - p_{JI}}{\tau} \ln \frac{p_{IJ}}{p_{JI}}
,\end{equation}
here rewritten in our notation. Note that, up to time constants, the semi-Markov entropy production is precisely the entropy production of the embedded Markov chain as a process in discrete time. The desired bound follows from \autoref{SM:tur1:proesmansI}  after bounding $\mathcal{N}$ appropriately from below. By introducing $x_{IJ} = \ln (p_{IJ}/p_{JI})$, we can write $\braket{\Delta S_{\text{tot}}} \tau/T = \sum_{IJ} p_{IJ}x_{IJ}$. Since the $p_{IJ}$ can be seen as a probability measure, we can use Jensen's inequality twice in the form
\begin{equation}
\frac{1}{1 + \exp \left( \braket{\Delta S_{\text{tot}}} \frac{\tau}{T} \right)} = \frac{1}{1 + \exp \left( \sum_{IJ} p_{IJ}x_{IJ} \right)} \leq \frac{1}{1 + \sum_{IJ} p_{IJ} \exp \left( x_{IJ} \right)} \leq \sum_{IJ} p_{IJ} \frac{1}{1 + \exp \left( x_{IJ} \right)} = \mathcal{N}
\end{equation}
as the functions $f(x) = e^x$ and $g(y) = 1/(1 + y)$ are convex in the appropriate domain, i.e., for $x \in \mathbb{R}$ and $y \geq 0$, respectively. Rearranging terms yields
\begin{equation}
\mathcal{I} = \frac{T}{\tau} \left( \frac{1}{2 \mathcal{N}} - 1 \right) \leq \frac{T}{2 \tau} \left( e^{\braket{\Delta S_{\text{tot}}} \frac{\tau}{T}} - 1 \right)
,\end{equation}
which together with \eqref{SM:tur1:cr} gives the semi-Markov TUR 
\begin{equation}
\frac{\Braket{\mathcal{J}} ^2}{T \text{ Var } \mathcal{J}} \leq \frac{1}{2 \tau} \left( e^{\braket{\Delta S_{\text{tot}}} \frac{\tau}{T}} - 1 \right)
,\end{equation}
as presented in the main text in inequality \eqref{III:EQ5_ineq}.

\subsection{Alternative motivation for \eqref{SM:tur1:proesmans}}
Starting from an ansatz of the form \eqref{SM:tur1:ansatz1}, one can use a simple heuristic argument to motivate \eqref{SM:tur1:proesmans}. We return to the first equality of equation
\eqref{SM:tur1:fisherAnsatz}, but incorporate the symmetry condition $q_{IJ} = q_{JI}$, which is needed to scale the currents correctly (cf. equation \eqref{SM:tur1:4}).

By minimizing the resulting Fisher information
\begin{equation}
\mathcal{I} = \frac{T}{\tau} \sum_{I<J}  \left( 1 - \frac{q_{IJ}}{p_{IJ}} \right)^2 p_{IJ} + \left( 1 - \frac{q_{IJ}}{p_{JI}} \right)^2 p_{JI}
\end{equation}
with respect to the $q_{IJ}$, we arrive at
\begin{equation}
q_{IJ} = \frac{p_{IJ}p_{JI}}{p_{IJ} + p_{JI}}
,\end{equation}
which, however, is unnormalized. After working a normalization constant $\mathcal{N}$ in, we arrive at the ansatz \eqref{SM:tur1:proesmans}.

\section{Proof of the Markov-based TUR}

\subsection{Methods and sketch of proof}
The second main result assumes that the semi-Markov process emerges from a lumping of states on an underlying Markovian structure. We will prove that this class of processes satisfies a TUR of the form
\begin{equation}
    \frac{\Braket{\mathcal{J}}^2}{\text{ Var } \mathcal{J}} \leq \frac{\Braket{\Delta S_{\text{tot}}}}{2}
\label{SM:tur3:statement} 
\end{equation}
for semi-Markov currents $\mathcal{J}$ in the steady state. As in the proof of the semi-Markov TUR, the Markov-based TUR is derived from the Cramér-Rao inequality \eqref{SM:tur1:cr0} by designing an appropriate virtual comparison dynamics. In contrast to the semi-Markov TUR, however, the perturbation affects the Markov path weight of the fundamental process prior to coarse-graining. Therefore, this section starts with a preliminary discussion of properties of continuous-time Markov jump processes.

The actual proof proceeds in two steps. First, we will assume that in any given semi-Markov state only one of underlying Markov states is connected to the remaining network. This bottleneck structure allows us to design a comparison dynamics which scales visible currents without affecting the interior of a coarse-grained state. In the second step, we will present a decimation scheme that reduces generic coarse-grained semi-Markov states to those where our argument of the first step applies.

\subsection{Preliminaries I: Fisher information}
For Markov processes, the Fisher information of the corresponding path weight is known (see, e.g., \cite{dechant2018} for the general case of time-dependent jump rates). For our steady-state setting, the special case of time-independent rates suffices. We assume that the modified processes for $\theta \neq 0$ relates to the original process at $\theta = 0$ by modifying the jump rates $k_{ij}$ and the initial state $p_i(0)$ according to
\begin{align}
k_{ij}^\theta & = k_{ij} e^{\theta a_{ij}} \nonumber, \\
p_i^\theta (0)& = p_i(0) + \theta \tilde{p}_i
\label{SM:tur2:ansatzGeneral}
.\end{align}
Then, the Fisher information evaluated at the physical process at $\theta = 0$ becomes
\begin{equation}
\mathcal{I} \equiv \lim_{\theta \to 0} \frac{2}{\theta^2} D_{\text{KL}}(\theta||0) 
= \sum_{ij} \int _0 ^T dt p_i(t) k_{ij} a_{ij}^2 + \sum _i \frac{\tilde{p}_i^2}{p_i(0)}
\label{SM:tur2:fisherGeneral}
.\end{equation} 
Note that initializing the physical system in the NESS results in $p_i(t) = p_i(0)$ and hence a trivial time integral. Usually, the derivation of a TUR proceeds by designing a comparison dynamics which scales all steady-state currents with the same factor, i.e.,  $\left. \partial _\theta \right| _{0} \Braket{\mathcal{J}}^{\theta} =  \Braket{\mathcal{J}}$ or, equivalently, $\left. \partial _\theta \right| _{0} j_{ij}^\theta = j_{ij}$. However, we proceed differently by allowing different scaling of currents at different links $i \leftrightarrow j$, written as
\begin{equation}
j_{ij}^\theta = j_{ij} + \theta \tilde{j}_{ij} 
\label{SM:tur2:scaling}
,\end{equation} 
in an infinitesimal region around $\theta = 0$ with parameters $\tilde{j}_{ij}$ yet to be specified. A choice of perturbed rates $k_{ij}^\theta$ which fulfills \eqref{SM:tur2:scaling} is obtained by substituting
\begin{equation}
a_{ij} = \frac{\tilde{j}_{ij} - \tilde{p}_i k_{ij} + \tilde{p}_j k_{ji}}{p_i(t) k_{ij} + p_j(t) k_{ji}}
\label{SM:tur2:scalingChoice}
\end{equation} 
into the ansatz \eqref{SM:tur2:ansatzGeneral}. Moreover, this choice of rates is optimal in the sense that any other choice consistent with \eqref{SM:tur2:scaling} results in a higher Fisher information (i.e., ultimately a looser TUR bound) than
\begin{equation}
\mathcal{I} = T \sum_{i < j} \frac{\left( \tilde{j}_{ij} - \tilde{p}_i k_{ij} + \tilde{p}_j k_{ji} \right)^2}{p_i(t) k_{ij} + p_j(t) k_{ji}} + \sum _i \frac{\tilde{p}_i^2}{p_i(0)}
\label{SM:tur2:fisherSimple}
,\end{equation} 
the Fisher information associated with the choice \eqref{SM:tur2:scalingChoice}, which is calculated from equation \eqref{SM:tur2:fisherGeneral}. One can understand the choice \eqref{SM:tur2:scalingChoice} as the result of an optimization over the activity at each link, as this quantity remains unspecified. This optimality is also discussed in \cite{shiraishi2021}.

\subsection{Preliminaries II: Cycles and cycle currents}
If the physical system is in a NESS, a sensible comparison dynamics for $\theta \neq 0$ can be chosen as time-independent, preferably with the same stationary distribution $p_i$. For $N$ states, this imposes $N - 1$ constraints on the possible scaling parameters $\tilde{j}_{ij}$ in equation \eqref{SM:tur2:scaling}. Consequently, there can be at most $E - N + 1$ independent choices if $E$ specifies the number of edges. By Euler's formula, this number equals the number of fundamental cycles present in the system. By scaling these fundamental cycles individually, we can formulate cycle-specific TUR-like relations as in \cite{polettini2021} by using jargon from network theory \cite{schnakenberg1976, andrieux2007, puglisi2010}.

\begin{figure}[htb]
\centering
\includegraphics[width=\linewidth]{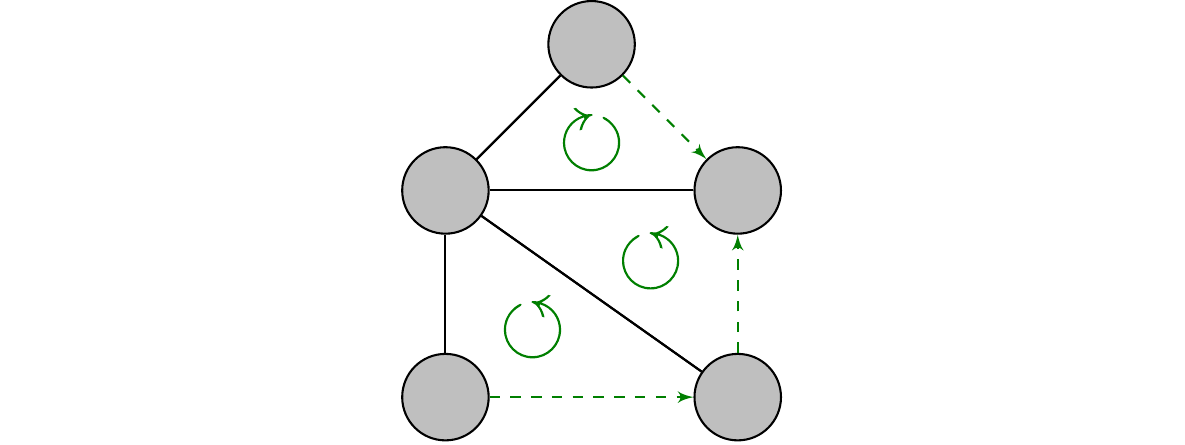}
\caption[cg1]{A graph which contains five states and three fundamental cycles. A possible spanning tree is indicated with solid, black lines. Any of the three fundamental cycles contains one \emph{chord}, i.e. one edge which is not contained in the spanning tree. The dashed green arrow indicates an orientation of the chord and its corresponding fundamental cycle.}
\label{SM:fig:cyclegtraph} 
\end{figure}

We make use of a decomposition into fundamental cycles. Informally, these cycles can be understood as a basis for all possible cycles in the network. The construction of fundamental cycles requires the identification of a spanning tree as a subset of the graph underlying a Markov state network. This spanning tree is obtained by connecting all $N$ states with $N-1$ edges from the original graph such that a simply connected substructure is obtained. An example of a spanning tree is depicted in black in \autoref{SM:fig:cyclegtraph}.

The remaining $E - N + 1$ edges which are not included in the spanning tree are commonly referred to as chords. Any of these chords give rise to a single fundamental cycle, the only closed cycle in the subgraph consisting of the spanning tree and a single chord. By construction, any cycle can be decomposed into a sum of fundamental cycles. Moreover, a chosen orientation for the chord induces one on the corresponding fundamental cycle. In the example in \autoref{SM:fig:cyclegtraph} the chords and their respective fundamental cycles are depicted as green arrows.

Given a cycle $\mathcal{C}_\alpha$, we define its directed adjacency matrix as
\begin{equation}\chi ^\alpha _{ij} = 
\begin{cases}
1, & i \to j \text{ contained in } \mathcal{C}_\alpha \\
-1, & j \to i \text{ contained in } \mathcal{C}_\alpha \\
0, & \text{otherwise.}
\end{cases}
.\label{SM:tur2:multiChi}\end{equation} If the link $i \to j$ is the chord of a fundamental cycle $\mathcal{C}_\beta$, then
\begin{equation}
\chi ^\alpha _{ij} = \delta _{\alpha \beta}
,\end{equation} as $\alpha$ runs through the set of fundamental cycles $\{ \mathcal{C}_\alpha \}$. This gives rise to the definition of a trajectory-dependent stochastic current $\mathcal{J}_\alpha$ by counting jumps along the chord, i.e.,
\begin{equation}
\label{SM:tur2:cycleCurrDef}
\mathcal{J}_\alpha \equiv \frac{1}{T} \left( n_{i \to j} - n_{j \to i} \right)
.\end{equation} As a consequence, 
\begin{equation}
j_\alpha \equiv \braket{\mathcal{J}_\alpha} = j_{ij}
\label{SM:tur2:cycleCurrAvg}
\end{equation} holds true for the expectation values. The decomposition into fundamental cycles can be transferred to steady state currents and entropy production as well. First, we define the cycle affinity 
\begin{equation}
\mathcal{A}_\alpha \equiv \ln \left( \prod_{i \rightarrow j \text{ in } \mathcal{C}_\alpha} \frac{k_{ij}}{k_{ji}} \right) = \sum _{i < j} \chi ^\alpha _{ij} \ln \frac{k_{ij}}{k_{ji}} = \sum _{i < j} \chi ^\alpha _{ij} \ln \frac{p_i k_{ij}}{p_j k_{ji}}
,\end{equation} which is the entropy produced by completing the cycle once. We can then decompose transition currents and entropy production rate as 
\begin{eqnarray}
\label{SM:tur2:multiCurr}
j_{ij} & = \sum _\alpha \chi ^\alpha _{ij} j_\alpha \nonumber \\
\sigma & = \sum _\alpha j_\alpha \mathcal{A}_\alpha
,\end{eqnarray}  where $j_\alpha$ and $\mathcal{A}_\alpha$ are the steady-state currents and affinities of fundamental cycles $\mathcal{C}_\alpha$, respectively. This is a consequence of Kirchhoff's current law, since $0 = - \dot{p}_i = \sum _j j_{ij}$ in the steady state. 

Now, we return to the design of a comparison dynamics with a parameter $\theta$. On the level of averaged currents, we are free to scale all cycle currents $j_\alpha$ of a set of fundamental cycles $\{ \mathcal{C}_\alpha \}$ individually, e.g. written as
\begin{equation}
j_\alpha^\theta = (1 + c_\alpha \theta) j_\alpha
.\end{equation} For any choice of parameters $c_\alpha$, the comparison dynamics still respects Kirchhoff's law at each vertex. Using the decomposition \eqref{SM:tur2:multiCurr}, the current at link $i \to j$ is scaled according to
\begin{equation}
j_{ij} ^\theta = \sum _{\alpha} \chi ^\alpha _{ij} \left( 1 + c_\alpha \theta \right) j_\alpha
\label{SM:tur2:genericAnsatz}
.\end{equation} Comparing this equation to the generic scheme \eqref{SM:tur2:scaling} for comparison dynamics, we can use equation \eqref{SM:tur2:fisherSimple} to calculate the Fisher information term
\begin{equation}
\mathcal{I} = T \sum_{i < j} \frac{\left( \sum _{\alpha} \chi ^\alpha _{ij} c_\alpha j_\alpha \right)^2}{p_i k_{ij} + p_j k_{ji}}
\label{SM:tur2:genericFisher}
.\end{equation} Note that we have set $\tilde{p}_i = 0$, because the stationary state remains unchanged. In the next step of the proof, we will specify the parameters $c_\alpha$.

\subsection{Step 1: Systems with one entry/exit state}\label{SubS_Cite}
We use capitals $I, J, ...$ for semi-Markov states, which consist of Markov states $i, j, ...$ on a fundamental level. We assume that every Markov-state is mapped to a unique semi-Markov state, but semi-Markov states containing a single Markov state are allowed. In addition, we make use of the notation for the inner and outer boundary of a semi-Markov state $I$ (when viewed as a set of Markov states) which was adopted in the main paper and in \cite{Qian2007}:
\begin{itemize}
    \item We write $j \in \partial_{\text{out}}I$ if $j \notin I$ and there is a state $i \in I$ with $k_{ij} > 0$
    \item We write $i \in \partial_{\text{in}}I$ if $i \in I$ and there is a state $j \notin I$ with $k_{ji} > 0$ 
    \item The set of all inner boundaries is denoted $D_{\text{in}}$: $D_{\text{in}} \equiv \cup_{I} \partial_{\text{in}}I$.
\end{itemize}
Assuming that $k_{ij} > 0$ implies $k_{ji} > 0$, a state in $\partial_{\text{out}}I$ can be reached from a state within $I$ only by passing through a state $i \in \partial_{\text{in}}I$ at some point. For now, we assume that for all semi-Markov states $I$, exactly one Markov state is contained in the set $\partial_{\text{in}}I$ (see figure 2 in the main text for an example). We also assume that the Markov process is in the steady state and that the underlying Markov network is connected. 



\subsubsection*{Comparison dynamics and correspondence of Markov and semi-Markov current}
Let us denote the unique boundary states of $I$ and $J$ by corresponding lowercase letters, i.e., $i(I) \in \partial_{\text{in}}I$ or simply $i \in \partial_{\text{in}}I$ and $j \in \partial_{\text{in}}J$, respectively. Since cycles are self-avoiding paths, they can visit any state at most once. In particular, a cycle which visits a state $k \in I \setminus \partial_{\text{in}}I$ cannot leave $I$, because it would pass through $i(I)$ more than once. Therefore, any fundamental cycle $\mathcal{C}_{\alpha}$ is either completely hidden within a semi-Markov state, or connects only states in $D_{\text{in}}$ and therefore remains completely visible in the semi-Markov description. We construct a comparison dynamics which scales these visible cycles from ansatz \eqref{SM:tur2:genericAnsatz} with parameters $c_\alpha$ in the form
\begin{equation}
c_\alpha = \begin{cases}
0, & \text{there is a semi-Markov state } I \text{ with } \mathcal{C}_{\alpha} \subseteq I \\
1, & \text{otherwise}
\end{cases}
.\label{SM:tur3:parameters}\end{equation} This construction ensures
\begin{equation}j_{ij}^\theta = 
\begin{cases}
j_{ij} & i \text{ and }j \text{ are from the same semi-Markov state } \\
(1 + \theta) j_{ij} & i \text{ and }j \text{ are from different semi-Markov states }
\end{cases}
\label{SM:tur3:scaling1}\end{equation}
for the steady state currents. Note that  the link corresponding to a semi-Markov transition $I \to J$ on the Markovian level, namely $i(I) \to j(J)$, is unique. In particular, we have
\begin{equation}
j_{ij} = j_{i(I)j(J)} = j_{IJ} \text{ for } i \in \partial_{\text{in}}I, j \in \partial_{\text{in}}J
.\label{SM:tur3:currentsMatch2}\end{equation}
This argument applies to stochastic semi-Markov currents as well, as jumps of the form $I \to J$ can be traced back to $i(I) \to j(J)$ without ambiguities. Thus, an empirical current $\mathcal{J}$ described by antisymmetric increments $q_{IJ}$ has a unique equivalent on the Markov level given by
\begin{equation}
    \mathcal{J} = \frac{1}{T}\sum_{IJ}n_{I\rightarrow J}q_{IJ} = \frac{1}{T}\sum_{I J}n_{i(I)\rightarrow j(J)}q_{i(I),j(J)}
.\label{SM:tur3:currentsMatch}\end{equation}
In view of a scaling condition to prove the TUR, we note that combining the equations \eqref{SM:tur3:scaling1}, \eqref{SM:tur3:currentsMatch2} and \eqref{SM:tur3:currentsMatch} implies
\begin{equation}
    \left. \partial _\theta \right| _{0} \Braket{\mathcal{J}}^{\theta} =  \Braket{\mathcal{J}}
\label{SM:tur3:scalingCurr}
\end{equation} 
for all semi-Markov currents $\mathcal{J}$.

\subsubsection*{Relation between semi-Markov and Markov steady state entropy production rate}
To utilize a TUR argument on the underlying Markovian level, we have to reformulate the semi-Markov entropy production [right-hand side of equation \eqref{SM:tur3:statement}]. Based on the defining equation of a semi-Markov steady state,  $0 = - \partial _t \pi_I = \sum _J j_{I J}$, we can derive
\begin{align}
    \sum _{I<J} j_{I J} \ln \frac{f(I)}{f(J)} = \sum _{I<J} j_{I J} \ln f(I) - \sum_{J < I} j_{J I} \ln f(I)
    = \sum _{I \neq J} j_{I J} \ln f(I) = \sum _I \left( \sum _J j_{I J} \right) \ln f(I) = 0
\label{SM:tur3:lemma}\end{align}
for arbitrary state functions $f(I)$. Setting $f(I) = \pi_I/(\braket{t_{I}} \sum _{k \notin I)} k_{i(I)k})$, we can split this term off the semi-Markov entropy to obtain
\begin{equation}
    \braket{\Delta S_{\text{tot}}} = T \sum _{I<J} j_{I J} \ln \frac{\pi_{I}\cdot p_{I\to J}/\braket{t_{I}}}{\pi_{J}\cdot p_{J\to I}/\braket{t_{J}}}
    = T \sum _{I<J} j_{I J} \ln \frac{p_{I\to J} (\sum_{k \notin I} k_{ik})}{p_{J\to I} (\sum _{k \notin J} k_{jk})} = T \sum _{i(I)<j(J)} j_{i j} \ln \frac{k_{ij}}{k_{ji}}
.\end{equation} In the last line, we have used \eqref{SM:tur3:currentsMatch2} and 
\begin{equation}
    p_{I\to J} = p(\text{jump } i \to j | \text{jump } i \text{ to state } k \in \partial_{\text{out}}I) = k_{ij}/\sum _{k \notin I}  k_{ik},
\end{equation} 
because the system must jump along one of the links $i \to k \notin I$ if we observe a jump out of $I$. In other words, the semi-Markov steady state entropy production rate is precisely the Markov entropy production along the visible links,
\begin{align}
    \frac{\braket{\Delta S_{\text{tot}}}}{T} = \sum _{\substack{i<j \\ i,j \in D_{\text{in}}}} j_{i j} \ln \frac{p_i k_{ij}}{p_j k_{ji}}
\label{SM:tur3:entropy}
\end{align} where the steady state probabilities $p_i$ of the Markov process can be included by essentially the same argument as equation \eqref{SM:tur3:lemma}.

\subsubsection*{Bounding Fisher information against entropy}
Our comparison dynamics follows the ansatz \eqref{SM:tur2:genericAnsatz} with the parameters $c_\alpha$ specified in \eqref{SM:tur3:parameters}. Thus, the corresponding Fisher information term, equation
\eqref{SM:tur2:genericFisher}, can be recast as
\begin{equation}
\mathcal{I} = T \sum_{i < j} \frac{\left( \sum _{\alpha} \chi ^\alpha _{ij} c_\alpha j_\alpha \right)^2}{p_i k_{ij} + p_j k_{ji}} = T \sum_{\substack{i<j \\ i,j \in D_{\text{in}}}} \frac{j_{ij}^2}{p_i k_{ij} + p_j k_{ji}}
,\end{equation}
because for any ''hidden`` link $i \to j$ within a semi-Markov state $I$, all fundamental cycles $\mathcal{C}_\alpha$ satisfy either $\chi^\alpha_{ij} = 0$ or $c_\alpha = 0$, i.e., they do not pass through $i \to j$ or are not scaled . Similarly, cycles through a visible link $i \to j$ with $i, j \in D_{\text{in}}$ satisfy $c_\alpha = 1$ and therefore recover the full current $j_{ij}$ [cf. equations \eqref{SM:tur3:parameters} and \eqref{SM:tur3:scaling1}]. The last step,
\begin{equation}
\mathcal{I} = T \sum_{\substack{i<j \\ i,j \in D_{\text{in}}}} j_{ij} \frac{p_i k_{ij} - p_j k_{ji}}{p_i k_{ij} + p_j k_{ji}} \leq \frac{T}{2} \sum _{\substack{i<j \\ i,j \in D_{\text{in}}}} j_{i j} \ln \frac{p_i k_{ij}}{p_j k_{ji}}
= \frac{1}{2} \braket{\Delta S_{\text{tot}}}
,\label{SM:tur3:rhs} \end{equation} follows from $[\exp(x) - 1]/[\exp(x) + 1] = \tanh (x/2) \leq x/2$ for $x = p_i k_{ij}/(p_j k_{ji}) \geq 0$ (or its reverse for $x < 0$). The final result is evident after plugging the equations \eqref{SM:tur3:scalingCurr} and \eqref{SM:tur3:entropy} into the Cramér-Rao inequality, resulting in
\begin{equation}
\Braket{\mathcal{J}} ^2 = \left( \left. \partial _\theta \right| _{0} \Braket{\mathcal{J}}^{\theta} \right)^2 \leq \mathcal{I} \text{ Var } \mathcal{J} \leq \frac{\braket{\Delta S_{\text{tot}}}}{2} \text{Var } \mathcal{J}
\label{SM:tur3:cr} 
,\end{equation}  
the result referred to as the Markov-based TUR in the main text.

\subsection{Step 2: Systems with more than one entry/exit state}
So far, the result \eqref{SM:tur3:cr} has been established for systems with only one state in $\partial_{\text{in}}I$. To reduce general Markov networks to the case considered in the previous section, we will present a decimation scheme which alters the underlying Markov network, but not the semi-Markov process itself. In other words, the semi-Markov kernel $\psi_{I \to J}(t)$, which characterizes a semi-Markov process, is kept invariant. 

\begin{figure}[htb]
\includegraphics[width=\linewidth]{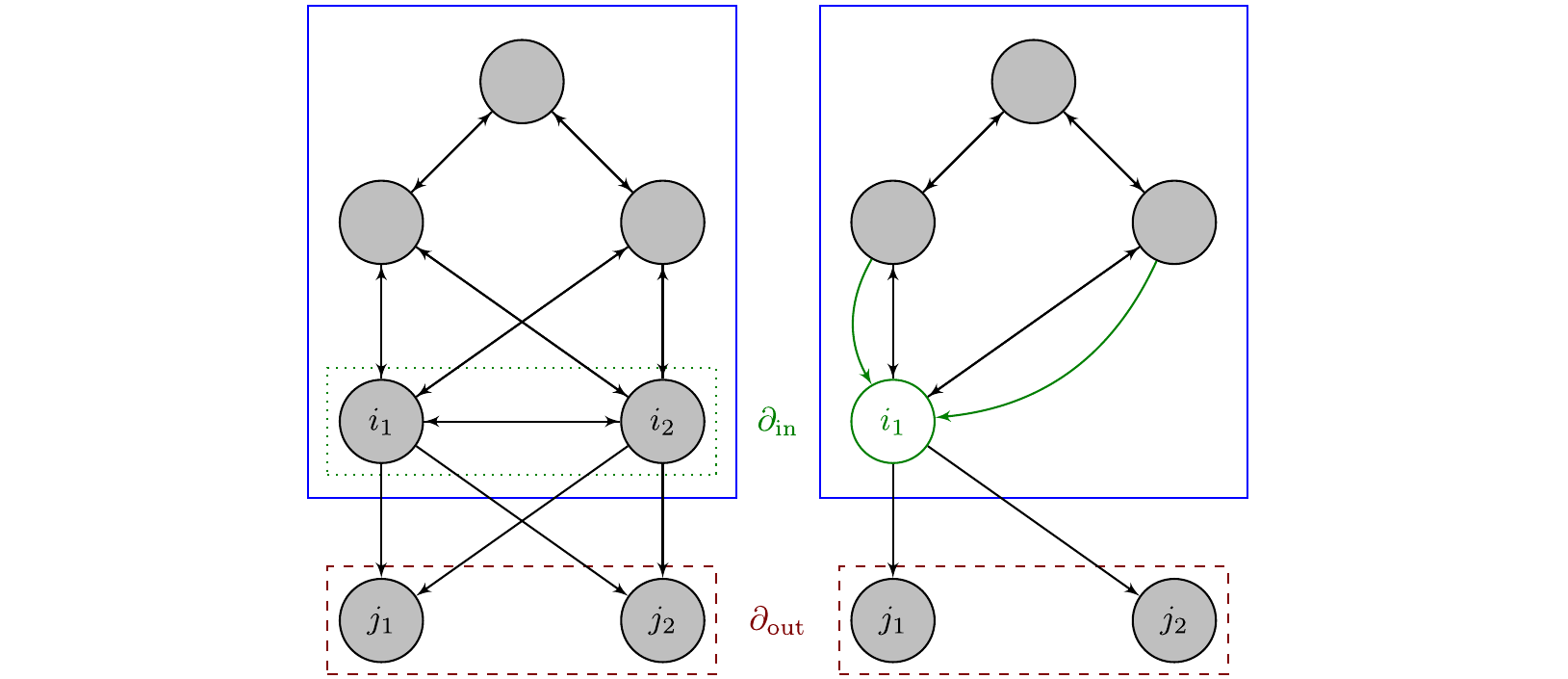}
\caption[cg2]{\textit{Left:} The escape problem associated with $\psi_{I \to J}(t)$ treats states in $ \partial_{\text{out}} I$ as absorbing states. On the Markov level, the escape problem corresponding to $\psi_{I \to J}(t)$ is given by $\widetilde{\psi}_{i \to j}(t)$ with $i \in \partial_{\text{in}} I$ and $j \in \partial_{\text{out}} I$. \textit{Right:} The decimation scheme for Markov states redirects links into $\partial_{\text{in}} I$ to a single boundary state $i_1, \in \partial_{\text{in}} I$ in step 3. In step 4, the state $i_2$ is removed.}
\label{SM:tur4:decimation} 
\end{figure}

In our description of the semi-Markov process as a coarse-grained Markov network obtained by grouping states, $\psi_{I \to J}(t)$ is the probability to spend the time $t$ in states $i \in I$ before leaving to a state $j \in \partial_{\text{out}} I$. Since $I$ is entered at states $i \in \partial_{\text{in}}I$, calculating $\psi_{I \to J}(t)$ amounts to solving the problem of escaping to (absorbing) states $j \in \partial_{\text{out}} I$ from a starting point in $i \in \partial_{\text{in}}I$, as discussed in the main text and \cite{Qian2007}. As in the main text, let
\begin{equation}
    \widetilde{\psi}_{i \to j}(t) = p(\text{arrive in } j \in \partial_{\text{out}} I \text{ at } t | \text{start in } i \in \partial_{\text{in}} I \text{ at } t = 0)
\end{equation}
denote joint probability distribution of ending in state $j$ at time $t$ given $i$. The compound state $I$ can be considered a semi-Markov state if  $\widetilde{\psi}_{i \to j}(t)$ is independent of $i \in \partial_{\text{in}} I$ (cf. \cite{Qian2007}, Sec. III C). Note that the converse is not true in general. By choosing sufficiently symmetric incoming rates, e.g., $k_{ji_1} = k_{ji_2}$ for all $j \in \partial_{\text{out}} I$ in the left network in \autoref{SM:tur4:decimation}, knowing the past does not yield new information about $\psi_{I \to J}(t)$. Thus, even if $\widetilde{\psi}_{i_1 \to j}(t) \neq \widetilde{\psi}_{i_2 \to j}(t)$, the compound state $I$ can be semi-Markov ''by accident``.

If $\widetilde{\psi}_{i \to j} (t)$ is independent of $i$, it was proven that \cite{Qian2007}:
\begin{itemize}
    \item Direction-time independence is granted
    \item Every $i \in \partial_{\text{in}} I$ is connected to every state $j \in \partial_{\text{out}} I$ on the outer boundary. Moreover, the jump rate $k_{ij}$ is independent of $i$.
\end{itemize}
We can now describe the decimation scheme of altering the structure of a semi-Markov compound state $I$ so that $I$ has a single state $i_0 \in \partial_{\text{in}} I$ on the inner boundary.
\begin{enumerate}
    \item Fix an arbitrary $i_0 \in \partial_{\text{in}} I$. 
    \item Redirect all \emph{incoming} links of the form $k_{ji}$ with $j \in \partial_{\text{out}} I, i \in \partial_{\text{in}} I$: Redirect $j \to i$ to $j \to i_0$, i.e., there is now a single incoming link from $j$ into the semi-Markov state $I$ with rate $k_{ji_0}^{\text{new}} \equiv \sum _{i \in \partial_{\text{in}} I} k_{ji}$. 
    \item Redirect all \emph{internal} links $k \to i$ from any $k \in I \setminus \partial_{\text{in}} I$ to some $i \in \partial_{\text{in}}I$ to $i_0$ instead of $i$. Remove all links of the form $k_{i_1 i_2}$ with $i_1, i_2 \in \partial_{\text{in}} I$.
    \item Since there are no incoming rates into states $i \in \partial_{\text{in}} I$ other than $i_0$, all these states can be removed from the network without changing the semi-Markov path weight. These probability of being in one of these states is zero in the steady state.
    \item $i_0$ is now the only state left in $\partial_{\text{in}} I$. 
\end{enumerate}

It is crucial that $\widetilde{\psi}_{i \to j}(t)$  is independent of $i \in \partial_{\text{in}} I$, thus the steps $2$ and $3$ do not affect $\psi_{I \to J}(t)$. Intuitively, the escape problem is effectively reset upon (re)visiting $\partial_{\text{in}} I$, which allows us to choose which state $i \in \partial_{\text{in}} I$ is actually visited. After repeating the decimation scheme for all semi-Markov states $I$, the semi-Markov process is of the type considered in Appendix \ref{SubS_Cite}. In particular, \autoref{SM:tur3:cr},  
\begin{equation}
\Braket{\mathcal{J}} ^2 \leq \frac{\braket{\Delta S_{\text{tot}}}}{2} \text{Var } \mathcal{J}
\end{equation}  
remains valid. Note that subtle issues regarding the steady state of the underlying  Markov process are circumvented by the fundamental assumption that the coarse-grained dynamics is described by a semi-Markov process. The semi-Markov steady state is a function of the semi-Markov path weight, which remains unchanged. Therefore both the steady state of the original and modified underlying Markov process reproduce the semi-Markov steady state. 

\twocolumngrid
\clearpage

%


\begin{thebibliography}{54}%
\makeatletter
\providecommand \@ifxundefined [1]{%
 \@ifx{#1\undefined}
}%
\providecommand \@ifnum [1]{%
 \ifnum #1\expandafter \@firstoftwo
 \else \expandafter \@secondoftwo
 \fi
}%
\providecommand \@ifx [1]{%
 \ifx #1\expandafter \@firstoftwo
 \else \expandafter \@secondoftwo
 \fi
}%
\providecommand \natexlab [1]{#1}%
\providecommand \enquote  [1]{``#1''}%
\providecommand \bibnamefont  [1]{#1}%
\providecommand \bibfnamefont [1]{#1}%
\providecommand \citenamefont [1]{#1}%
\providecommand \href@noop [0]{\@secondoftwo}%
\providecommand \href [0]{\begingroup \@sanitize@url \@href}%
\providecommand \@href[1]{\@@startlink{#1}\@@href}%
\providecommand \@@href[1]{\endgroup#1\@@endlink}%
\providecommand \@sanitize@url [0]{\catcode `\\12\catcode `\$12\catcode
  `\&12\catcode `\#12\catcode `\^12\catcode `\_12\catcode `\%12\relax}%
\providecommand \@@startlink[1]{}%
\providecommand \@@endlink[0]{}%
\providecommand \url  [0]{\begingroup\@sanitize@url \@url }%
\providecommand \@url [1]{\endgroup\@href {#1}{\urlprefix }}%
\providecommand \urlprefix  [0]{URL }%
\providecommand \Eprint [0]{\href }%
\providecommand \doibase [0]{https://doi.org/}%
\providecommand \selectlanguage [0]{\@gobble}%
\providecommand \bibinfo  [0]{\@secondoftwo}%
\providecommand \bibfield  [0]{\@secondoftwo}%
\providecommand \translation [1]{[#1]}%
\providecommand \BibitemOpen [0]{}%
\providecommand \bibitemStop [0]{}%
\providecommand \bibitemNoStop [0]{.\EOS\space}%
\providecommand \EOS [0]{\spacefactor3000\relax}%
\providecommand \BibitemShut  [1]{\csname bibitem#1\endcsname}%
\let\auto@bib@innerbib\@empty
\bibitem [{\citenamefont {Montroll}\ and\ \citenamefont
  {Weiss}(1965)}]{montroll1965}%
  \BibitemOpen
  \bibfield  {author} {\bibinfo {author} {\bibfnamefont {E.~W.}\ \bibnamefont
  {Montroll}}\ and\ \bibinfo {author} {\bibfnamefont {G.~H.}\ \bibnamefont
  {Weiss}},\ }\href {https://doi.org/10.1063/1.1704269} {\bibfield  {journal}
  {\bibinfo  {journal} {J. Math. Phys.}\ }\textbf {\bibinfo {volume} {6}},\
  \bibinfo {pages} {167–181} (\bibinfo {year} {1965})}\BibitemShut {NoStop}%
\bibitem [{\citenamefont {Tunaley}(1974)}]{Tunaley1974}%
  \BibitemOpen
  \bibfield  {author} {\bibinfo {author} {\bibfnamefont {J.~K.~E.}\
  \bibnamefont {Tunaley}},\ }\href {https://doi.org/10.1007/bf01026731}
  {\bibfield  {journal} {\bibinfo  {journal} {J. Stat. Phys.}\ }\textbf
  {\bibinfo {volume} {11}},\ \bibinfo {pages} {397} (\bibinfo {year}
  {1974})}\BibitemShut {NoStop}%
\bibitem [{\citenamefont {Tunaley}(1975)}]{Tunaley1975}%
  \BibitemOpen
  \bibfield  {author} {\bibinfo {author} {\bibfnamefont {J.~K.~E.}\
  \bibnamefont {Tunaley}},\ }\href {https://doi.org/10.1007/bf01024180}
  {\bibfield  {journal} {\bibinfo  {journal} {J. Stat. Phys.}\ }\textbf
  {\bibinfo {volume} {12}},\ \bibinfo {pages} {1} (\bibinfo {year}
  {1975})}\BibitemShut {NoStop}%
\bibitem [{\citenamefont {Tunaley}(1976)}]{Tunaley1976}%
  \BibitemOpen
  \bibfield  {author} {\bibinfo {author} {\bibfnamefont {J.~K.~E.}\
  \bibnamefont {Tunaley}},\ }\href {https://doi.org/10.1007/bf01040704}
  {\bibfield  {journal} {\bibinfo  {journal} {J. Stat. Phys.}\ }\textbf
  {\bibinfo {volume} {14}},\ \bibinfo {pages} {461} (\bibinfo {year}
  {1976})}\BibitemShut {NoStop}%
\bibitem [{\citenamefont {Kutner}\ and\ \citenamefont
  {Masoliver}(2017)}]{Kutner2017_Rev}%
  \BibitemOpen
  \bibfield  {author} {\bibinfo {author} {\bibfnamefont {R.}~\bibnamefont
  {Kutner}}\ and\ \bibinfo {author} {\bibfnamefont {J.}~\bibnamefont
  {Masoliver}},\ }\href {https://doi.org/10.1140/epjb/e2016-70578-3} {\bibfield
   {journal} {\bibinfo  {journal} {Eur. Phys. J. B}\ }\textbf {\bibinfo
  {volume} {90}},\ \bibinfo {pages} {50} (\bibinfo {year} {2017})}\BibitemShut
  {NoStop}%
\bibitem [{\citenamefont {Masuda}, \citenamefont {Porter},\ and\ \citenamefont
  {Lambiotte}(2017)}]{Masuda2017_Rev}%
  \BibitemOpen
  \bibfield  {author} {\bibinfo {author} {\bibfnamefont {N.}~\bibnamefont
  {Masuda}}, \bibinfo {author} {\bibfnamefont {M.~A.}\ \bibnamefont {Porter}},\
  and\ \bibinfo {author} {\bibfnamefont {R.}~\bibnamefont {Lambiotte}},\ }\href
  {https://doi.org/https://doi.org/10.1016/j.physrep.2017.07.007} {\bibfield
  {journal} {\bibinfo  {journal} {Phys. Rep.}\ }\textbf {\bibinfo {volume}
  {716-717}},\ \bibinfo {pages} {1} (\bibinfo {year} {2017})}\BibitemShut
  {NoStop}%
\bibitem [{\citenamefont {Wang}\ and\ \citenamefont {Qian}(2007)}]{Qian2007}%
  \BibitemOpen
  \bibfield  {author} {\bibinfo {author} {\bibfnamefont {H.}~\bibnamefont
  {Wang}}\ and\ \bibinfo {author} {\bibfnamefont {H.}~\bibnamefont {Qian}},\
  }\href {https://doi.org/10.1063/1.2432065} {\bibfield  {journal} {\bibinfo
  {journal} {J. Math. Phys.}\ }\textbf {\bibinfo {volume} {48}},\ \bibinfo
  {pages} {013303} (\bibinfo {year} {2007})}\BibitemShut {NoStop}%
\bibitem [{\citenamefont {Martínez}\ \emph {et~al.}(2019)\citenamefont
  {Martínez}, \citenamefont {Bisker}, \citenamefont {Horowitz},\ and\
  \citenamefont {Parrondo}}]{martinez2019}%
  \BibitemOpen
  \bibfield  {author} {\bibinfo {author} {\bibfnamefont {I.~A.}\ \bibnamefont
  {Martínez}}, \bibinfo {author} {\bibfnamefont {G.}~\bibnamefont {Bisker}},
  \bibinfo {author} {\bibfnamefont {J.~M.}\ \bibnamefont {Horowitz}},\ and\
  \bibinfo {author} {\bibfnamefont {J.~M.~R.}\ \bibnamefont {Parrondo}},\
  }\href {https://doi.org/10.1038/s41467-019-11051-w} {\bibfield  {journal}
  {\bibinfo  {journal} {Nat. Comm.}\ }\textbf {\bibinfo {volume} {10}},\
  \bibinfo {pages} {3542} (\bibinfo {year} {2019})}\BibitemShut {NoStop}%
\bibitem [{\citenamefont {Teza}\ and\ \citenamefont {Stella}(2020)}]{Teza2020}%
  \BibitemOpen
  \bibfield  {author} {\bibinfo {author} {\bibfnamefont {G.}~\bibnamefont
  {Teza}}\ and\ \bibinfo {author} {\bibfnamefont {A.~L.}\ \bibnamefont
  {Stella}},\ }\href {https://doi.org/10.1103/PhysRevLett.125.110601}
  {\bibfield  {journal} {\bibinfo  {journal} {Phys. Rev. Lett.}\ }\textbf
  {\bibinfo {volume} {125}},\ \bibinfo {pages} {110601} (\bibinfo {year}
  {2020})}\BibitemShut {NoStop}%
\bibitem [{\citenamefont {Klafter}, \citenamefont {Blumen},\ and\ \citenamefont
  {Shlesinger}(1987)}]{klafter1987}%
  \BibitemOpen
  \bibfield  {author} {\bibinfo {author} {\bibfnamefont {J.}~\bibnamefont
  {Klafter}}, \bibinfo {author} {\bibfnamefont {A.}~\bibnamefont {Blumen}},\
  and\ \bibinfo {author} {\bibfnamefont {M.~F.}\ \bibnamefont {Shlesinger}},\
  }\href {https://doi.org/10.1103/PhysRevA.35.3081} {\bibfield  {journal}
  {\bibinfo  {journal} {Phys. Rev. A}\ }\textbf {\bibinfo {volume} {35}},\
  \bibinfo {pages} {3081} (\bibinfo {year} {1987})}\BibitemShut {NoStop}%
\bibitem [{\citenamefont {Bouchaud}\ and\ \citenamefont
  {Georges}(1990)}]{bouchaud1990}%
  \BibitemOpen
  \bibfield  {author} {\bibinfo {author} {\bibfnamefont {J.-P.}\ \bibnamefont
  {Bouchaud}}\ and\ \bibinfo {author} {\bibfnamefont {A.}~\bibnamefont
  {Georges}},\ }\href
  {https://www.sciencedirect.com/science/article/pii/037015739090099N}
  {\bibfield  {journal} {\bibinfo  {journal} {Phys. Rep.}\ }\textbf {\bibinfo
  {volume} {195}},\ \bibinfo {pages} {127} (\bibinfo {year}
  {1990})}\BibitemShut {NoStop}%
\bibitem [{\citenamefont {Meerschaert}\ and\ \citenamefont
  {Straka}(2014)}]{meerschaert2014}%
  \BibitemOpen
  \bibfield  {author} {\bibinfo {author} {\bibfnamefont {M.~M.}\ \bibnamefont
  {Meerschaert}}\ and\ \bibinfo {author} {\bibfnamefont {P.}~\bibnamefont
  {Straka}},\ }\href {https://doi.org/10.1214/13-AOP905} {\bibfield  {journal}
  {\bibinfo  {journal} {Ann. Probab.}\ }\textbf {\bibinfo {volume} {42}},\
  \bibinfo {pages} {1699} (\bibinfo {year} {2014})}\BibitemShut {NoStop}%
\bibitem [{\citenamefont {Metzler}\ \emph {et~al.}(2014)\citenamefont
  {Metzler}, \citenamefont {Jeon}, \citenamefont {Cherstvy},\ and\
  \citenamefont {Barkai}}]{Barkai2014_Rev}%
  \BibitemOpen
  \bibfield  {author} {\bibinfo {author} {\bibfnamefont {R.}~\bibnamefont
  {Metzler}}, \bibinfo {author} {\bibfnamefont {J.-H.}\ \bibnamefont {Jeon}},
  \bibinfo {author} {\bibfnamefont {A.~G.}\ \bibnamefont {Cherstvy}},\ and\
  \bibinfo {author} {\bibfnamefont {E.}~\bibnamefont {Barkai}},\ }\href
  {https://doi.org/10.1039/C4CP03465A} {\bibfield  {journal} {\bibinfo
  {journal} {Phys. Chem. Chem. Phys.}\ }\textbf {\bibinfo {volume} {16}},\
  \bibinfo {pages} {24128} (\bibinfo {year} {2014})}\BibitemShut {NoStop}%
\bibitem [{\citenamefont {Nelson}(1999)}]{Nelson1999}%
  \BibitemOpen
  \bibfield  {author} {\bibinfo {author} {\bibfnamefont {J.}~\bibnamefont
  {Nelson}},\ }\href {https://doi.org/10.1103/PhysRevB.59.15374} {\bibfield
  {journal} {\bibinfo  {journal} {Phys. Rev. B}\ }\textbf {\bibinfo {volume}
  {59}},\ \bibinfo {pages} {15374} (\bibinfo {year} {1999})}\BibitemShut
  {NoStop}%
\bibitem [{\citenamefont {Koch}\ and\ \citenamefont {Brady}(1988)}]{Koch1988}%
  \BibitemOpen
  \bibfield  {author} {\bibinfo {author} {\bibfnamefont {D.~L.}\ \bibnamefont
  {Koch}}\ and\ \bibinfo {author} {\bibfnamefont {J.~F.}\ \bibnamefont
  {Brady}},\ }\href {https://doi.org/10.1063/1.866716} {\bibfield  {journal}
  {\bibinfo  {journal} {Phys. Fluids}\ }\textbf {\bibinfo {volume} {31}},\
  \bibinfo {pages} {965} (\bibinfo {year} {1988})}\BibitemShut {NoStop}%
\bibitem [{\citenamefont {Cortis}\ \emph {et~al.}(2006)\citenamefont {Cortis},
  \citenamefont {Harter}, \citenamefont {Hou}, \citenamefont {Atwill},
  \citenamefont {Packman},\ and\ \citenamefont {Green}}]{Cortis2006}%
  \BibitemOpen
  \bibfield  {author} {\bibinfo {author} {\bibfnamefont {A.}~\bibnamefont
  {Cortis}}, \bibinfo {author} {\bibfnamefont {T.}~\bibnamefont {Harter}},
  \bibinfo {author} {\bibfnamefont {L.}~\bibnamefont {Hou}}, \bibinfo {author}
  {\bibfnamefont {E.~R.}\ \bibnamefont {Atwill}}, \bibinfo {author}
  {\bibfnamefont {A.~I.}\ \bibnamefont {Packman}},\ and\ \bibinfo {author}
  {\bibfnamefont {P.~G.}\ \bibnamefont {Green}},\ }\href
  {https://agupubs.onlinelibrary.wiley.com/doi/abs/10.1029/2006WR004897}
  {\bibfield  {journal} {\bibinfo  {journal} {Water Resour. Res.}\ }\textbf
  {\bibinfo {volume} {42}} (\bibinfo {year} {2006})}\BibitemShut {NoStop}%
\bibitem [{\citenamefont {Ferreira}, \citenamefont {Pena},\ and\ \citenamefont
  {Romanazzi}(2016)}]{Ferreira2016}%
  \BibitemOpen
  \bibfield  {author} {\bibinfo {author} {\bibfnamefont {J.}~\bibnamefont
  {Ferreira}}, \bibinfo {author} {\bibfnamefont {G.}~\bibnamefont {Pena}},\
  and\ \bibinfo {author} {\bibfnamefont {G.}~\bibnamefont {Romanazzi}},\ }\href
  {https://doi.org/https://doi.org/10.1016/j.apm.2015.09.034} {\bibfield
  {journal} {\bibinfo  {journal} {Appl. Math. Model.}\ }\textbf {\bibinfo
  {volume} {40}},\ \bibinfo {pages} {1850} (\bibinfo {year}
  {2016})}\BibitemShut {NoStop}%
\bibitem [{\citenamefont {Höfling}\ and\ \citenamefont
  {Franosch}(2013)}]{Hoefling2013_Rev}%
  \BibitemOpen
  \bibfield  {author} {\bibinfo {author} {\bibfnamefont {F.}~\bibnamefont
  {Höfling}}\ and\ \bibinfo {author} {\bibfnamefont {T.}~\bibnamefont
  {Franosch}},\ }\href {https://doi.org/10.1088/0034-4885/76/4/046602}
  {\bibfield  {journal} {\bibinfo  {journal} {Rep. Prog. Phys.}\ }\textbf
  {\bibinfo {volume} {76}},\ \bibinfo {pages} {046602} (\bibinfo {year}
  {2013})}\BibitemShut {NoStop}%
\bibitem [{\citenamefont {Jeon}\ \emph {et~al.}(2011)\citenamefont {Jeon},
  \citenamefont {Tejedor}, \citenamefont {Burov}, \citenamefont {Barkai},
  \citenamefont {Selhuber-Unkel}, \citenamefont {Berg-S\o{}rensen},
  \citenamefont {Oddershede},\ and\ \citenamefont {Metzler}}]{Metzler_Exp2011}%
  \BibitemOpen
  \bibfield  {author} {\bibinfo {author} {\bibfnamefont {J.-H.}\ \bibnamefont
  {Jeon}}, \bibinfo {author} {\bibfnamefont {V.}~\bibnamefont {Tejedor}},
  \bibinfo {author} {\bibfnamefont {S.}~\bibnamefont {Burov}}, \bibinfo
  {author} {\bibfnamefont {E.}~\bibnamefont {Barkai}}, \bibinfo {author}
  {\bibfnamefont {C.}~\bibnamefont {Selhuber-Unkel}}, \bibinfo {author}
  {\bibfnamefont {K.}~\bibnamefont {Berg-S\o{}rensen}}, \bibinfo {author}
  {\bibfnamefont {L.}~\bibnamefont {Oddershede}},\ and\ \bibinfo {author}
  {\bibfnamefont {R.}~\bibnamefont {Metzler}},\ }\href
  {https://doi.org/10.1103/PhysRevLett.106.048103} {\bibfield  {journal}
  {\bibinfo  {journal} {Phys. Rev. Lett.}\ }\textbf {\bibinfo {volume} {106}},\
  \bibinfo {pages} {048103} (\bibinfo {year} {2011})}\BibitemShut {NoStop}%
\bibitem [{\citenamefont {Goiko}, \citenamefont {Bruyn},\ and\ \citenamefont
  {Heit}(2018)}]{Goiko2018}%
  \BibitemOpen
  \bibfield  {author} {\bibinfo {author} {\bibfnamefont {M.}~\bibnamefont
  {Goiko}}, \bibinfo {author} {\bibfnamefont {J.}~\bibnamefont {Bruyn}},\ and\
  \bibinfo {author} {\bibfnamefont {B.}~\bibnamefont {Heit}},\ }\href
  {https://doi.org/10.1016/j.bpj.2018.04.024} {\bibfield  {journal} {\bibinfo
  {journal} {Biophys. J.}\ }\textbf {\bibinfo {volume} {114}},\ \bibinfo
  {pages} {2887} (\bibinfo {year} {2018})}\BibitemShut {NoStop}%
\bibitem [{\citenamefont {Fabens}(1961)}]{Fabens1961}%
  \BibitemOpen
  \bibfield  {author} {\bibinfo {author} {\bibfnamefont {A.~J.}\ \bibnamefont
  {Fabens}},\ }\href
  {https://doi.org/https://doi.org/10.1111/j.2517-6161.1961.tb00395.x}
  {\bibfield  {journal} {\bibinfo  {journal} {J. R. Stat. Soc. Series B Stat.
  Methodol.}\ }\textbf {\bibinfo {volume} {23}},\ \bibinfo {pages} {113}
  (\bibinfo {year} {1961})}\BibitemShut {NoStop}%
\bibitem [{\citenamefont {Korolyuk}, \citenamefont {Brodi},\ and\ \citenamefont
  {Turbin}(1975)}]{Korolyuk1975}%
  \BibitemOpen
  \bibfield  {author} {\bibinfo {author} {\bibfnamefont {V.~S.}\ \bibnamefont
  {Korolyuk}}, \bibinfo {author} {\bibfnamefont {S.~M.}\ \bibnamefont
  {Brodi}},\ and\ \bibinfo {author} {\bibfnamefont {A.~F.}\ \bibnamefont
  {Turbin}},\ }\href {https://doi.org/10.1007/bf01097184} {\bibfield  {journal}
  {\bibinfo  {journal} {J. Sov. Math.}\ }\textbf {\bibinfo {volume} {4}},\
  \bibinfo {pages} {244} (\bibinfo {year} {1975})}\BibitemShut {NoStop}%
\bibitem [{\citenamefont {Janssen}(2013)}]{janssen2013}%
  \BibitemOpen
  \bibfield  {author} {\bibinfo {author} {\bibfnamefont {J.}~\bibnamefont
  {Janssen}},\ }\href@noop {} {\emph {\bibinfo {title} {Semi-Markov Models:
  Theory and Applications}}}\ (\bibinfo  {publisher} {Springer Science \&
  Business Media},\ \bibinfo {address} {Berlin Heidelberg},\ \bibinfo {year}
  {2013})\BibitemShut {NoStop}%
\bibitem [{\citenamefont {Scalas}(2006)}]{Scalas_2006}%
  \BibitemOpen
  \bibfield  {author} {\bibinfo {author} {\bibfnamefont {E.}~\bibnamefont
  {Scalas}},\ }\href
  {https://doi.org/https://doi.org/10.1016/j.physa.2005.11.024} {\bibfield
  {journal} {\bibinfo  {journal} {Phys. A: Stat. Mech. Appl.}\ }\textbf
  {\bibinfo {volume} {362}},\ \bibinfo {pages} {225} (\bibinfo {year}
  {2006})}\BibitemShut {NoStop}%
\bibitem [{\citenamefont {Seifert}(2012)}]{Seifert2012}%
  \BibitemOpen
  \bibfield  {author} {\bibinfo {author} {\bibfnamefont {U.}~\bibnamefont
  {Seifert}},\ }\href {https://doi.org/10.1088/0034-4885/75/12/126001}
  {\bibfield  {journal} {\bibinfo  {journal} {Rep. Prog. Phys.}\ }\textbf
  {\bibinfo {volume} {75}},\ \bibinfo {pages} {126001} (\bibinfo {year}
  {2012})}\BibitemShut {NoStop}%
\bibitem [{\citenamefont {Chari}(1994)}]{chari1994}%
  \BibitemOpen
  \bibfield  {author} {\bibinfo {author} {\bibfnamefont {M.~K.}\ \bibnamefont
  {Chari}},\ }\href {https://doi.org/10.1016/0167-6377(94)90051-5} {\bibfield
  {journal} {\bibinfo  {journal} {Oper. Res. Lett.}\ }\textbf {\bibinfo
  {volume} {15}},\ \bibinfo {pages} {157} (\bibinfo {year} {1994})}\BibitemShut
  {NoStop}%
\bibitem [{\citenamefont {Girardin}\ and\ \citenamefont
  {Limnios}(2003)}]{Girardin2003}%
  \BibitemOpen
  \bibfield  {author} {\bibinfo {author} {\bibfnamefont {V.}~\bibnamefont
  {Girardin}}\ and\ \bibinfo {author} {\bibfnamefont {N.}~\bibnamefont
  {Limnios}},\ }\href {http://www.jstor.org/stable/3216060} {\bibfield
  {journal} {\bibinfo  {journal} {J. Appl. Probab.}\ }\textbf {\bibinfo
  {volume} {40}},\ \bibinfo {pages} {1060} (\bibinfo {year}
  {2003})}\BibitemShut {NoStop}%
\bibitem [{\citenamefont {Esposito}\ and\ \citenamefont
  {Lindenberg}(2008)}]{Esposito2008}%
  \BibitemOpen
  \bibfield  {author} {\bibinfo {author} {\bibfnamefont {M.}~\bibnamefont
  {Esposito}}\ and\ \bibinfo {author} {\bibfnamefont {K.}~\bibnamefont
  {Lindenberg}},\ }\href {https://doi.org/10.1103/PhysRevE.77.051119}
  {\bibfield  {journal} {\bibinfo  {journal} {Phys. Rev. E}\ }\textbf {\bibinfo
  {volume} {77}},\ \bibinfo {pages} {051119} (\bibinfo {year}
  {2008})}\BibitemShut {NoStop}%
\bibitem [{\citenamefont {Andrieux}\ and\ \citenamefont
  {Gaspard}(2008)}]{andrieux2008}%
  \BibitemOpen
  \bibfield  {author} {\bibinfo {author} {\bibfnamefont {D.}~\bibnamefont
  {Andrieux}}\ and\ \bibinfo {author} {\bibfnamefont {P.}~\bibnamefont
  {Gaspard}},\ }\href {https://doi.org/10.1088/1742-5468/2008/11/P11007}
  {\bibfield  {journal} {\bibinfo  {journal} {J. Stat. Mech.}\ }\textbf
  {\bibinfo {volume} {2008}},\ \bibinfo {pages} {P11007} (\bibinfo {year}
  {2008})}\BibitemShut {NoStop}%
\bibitem [{\citenamefont {Maes}, \citenamefont {Netočný},\ and\ \citenamefont
  {Wynants}(2009)}]{Maes2009}%
  \BibitemOpen
  \bibfield  {author} {\bibinfo {author} {\bibfnamefont {C.}~\bibnamefont
  {Maes}}, \bibinfo {author} {\bibfnamefont {K.}~\bibnamefont {Netočný}},\
  and\ \bibinfo {author} {\bibfnamefont {B.}~\bibnamefont {Wynants}},\ }\href
  {https://doi.org/10.1088/1751-8113/42/36/365002} {\bibfield  {journal}
  {\bibinfo  {journal} {J. Phys. A}\ }\textbf {\bibinfo {volume} {42}},\
  \bibinfo {pages} {365002} (\bibinfo {year} {2009})}\BibitemShut {NoStop}%
\bibitem [{\citenamefont {Barato}\ and\ \citenamefont
  {Seifert}(2015)}]{barato2015}%
  \BibitemOpen
  \bibfield  {author} {\bibinfo {author} {\bibfnamefont {A.~C.}\ \bibnamefont
  {Barato}}\ and\ \bibinfo {author} {\bibfnamefont {U.}~\bibnamefont
  {Seifert}},\ }\href {https://doi.org/10.1103/PhysRevLett.114.158101}
  {\bibfield  {journal} {\bibinfo  {journal} {Phys. Rev. Lett.}\ }\textbf
  {\bibinfo {volume} {114}},\ \bibinfo {pages} {158101} (\bibinfo {year}
  {2015})}\BibitemShut {NoStop}%
\bibitem [{\citenamefont {Gingrich}\ \emph {et~al.}(2016)\citenamefont
  {Gingrich}, \citenamefont {Horowitz}, \citenamefont {Perunov},\ and\
  \citenamefont {England}}]{gingrich2016}%
  \BibitemOpen
  \bibfield  {author} {\bibinfo {author} {\bibfnamefont {T.~R.}\ \bibnamefont
  {Gingrich}}, \bibinfo {author} {\bibfnamefont {J.~M.}\ \bibnamefont
  {Horowitz}}, \bibinfo {author} {\bibfnamefont {N.}~\bibnamefont {Perunov}},\
  and\ \bibinfo {author} {\bibfnamefont {J.~L.}\ \bibnamefont {England}},\
  }\href {https://doi.org/10.1103/PhysRevLett.116.120601} {\bibfield  {journal}
  {\bibinfo  {journal} {Phys. Rev. Lett.}\ }\textbf {\bibinfo {volume} {116}},\
  \bibinfo {pages} {120601} (\bibinfo {year} {2016})}\BibitemShut {NoStop}%
\bibitem [{\citenamefont {Seifert}(2019)}]{seifert2019}%
  \BibitemOpen
  \bibfield  {author} {\bibinfo {author} {\bibfnamefont {U.}~\bibnamefont
  {Seifert}},\ }\href
  {https://doi.org/10.1146/annurev-conmatphys-031218-013554} {\bibfield
  {journal} {\bibinfo  {journal} {Annu. Rev. Condens. Matter Phys.}\ }\textbf
  {\bibinfo {volume} {10}},\ \bibinfo {pages} {171} (\bibinfo {year}
  {2019})}\BibitemShut {NoStop}%
\bibitem [{\citenamefont {Horowitz}\ and\ \citenamefont
  {Gingrich}(2020)}]{Horowitz2020}%
  \BibitemOpen
  \bibfield  {author} {\bibinfo {author} {\bibfnamefont {J.~M.}\ \bibnamefont
  {Horowitz}}\ and\ \bibinfo {author} {\bibfnamefont {T.~R.}\ \bibnamefont
  {Gingrich}},\ }\href {https://doi.org/10.1038/s41567-020-0853-5} {\bibfield
  {journal} {\bibinfo  {journal} {Nature Phys.}\ }\textbf {\bibinfo {volume}
  {16}},\ \bibinfo {pages} {599} (\bibinfo {year} {2020})}\BibitemShut
  {NoStop}%
\bibitem [{\citenamefont {Vu}\ and\ \citenamefont {Hasegawa}(2020)}]{vu2020}%
  \BibitemOpen
  \bibfield  {author} {\bibinfo {author} {\bibfnamefont {T.~V.}\ \bibnamefont
  {Vu}}\ and\ \bibinfo {author} {\bibfnamefont {Y.}~\bibnamefont {Hasegawa}},\
  }\href {https://doi.org/10.1088/1742-6596/1593/1/012006} {\bibfield
  {journal} {\bibinfo  {journal} {J. Phys. Conf. Ser.}\ }\textbf {\bibinfo
  {volume} {1593}},\ \bibinfo {pages} {012006} (\bibinfo {year}
  {2020})}\BibitemShut {NoStop}%
\bibitem [{\citenamefont {Seifert}(2005)}]{Seifert2005}%
  \BibitemOpen
  \bibfield  {author} {\bibinfo {author} {\bibfnamefont {U.}~\bibnamefont
  {Seifert}},\ }\href {https://doi.org/10.1103/PhysRevLett.95.040602}
  {\bibfield  {journal} {\bibinfo  {journal} {Phys. Rev. Lett.}\ }\textbf
  {\bibinfo {volume} {95}},\ \bibinfo {pages} {040602} (\bibinfo {year}
  {2005})}\BibitemShut {NoStop}%
\bibitem [{\citenamefont {Golding}\ and\ \citenamefont
  {Cox}(2006)}]{Golding2006}%
  \BibitemOpen
  \bibfield  {author} {\bibinfo {author} {\bibfnamefont {I.}~\bibnamefont
  {Golding}}\ and\ \bibinfo {author} {\bibfnamefont {E.~C.}\ \bibnamefont
  {Cox}},\ }\href {https://doi.org/10.1103/PhysRevLett.96.098102} {\bibfield
  {journal} {\bibinfo  {journal} {Phys. Rev. Lett.}\ }\textbf {\bibinfo
  {volume} {96}},\ \bibinfo {pages} {098102} (\bibinfo {year}
  {2006})}\BibitemShut {NoStop}%
\bibitem [{\citenamefont {Dix}\ and\ \citenamefont {Verkman}(2008)}]{Dix2008}%
  \BibitemOpen
  \bibfield  {author} {\bibinfo {author} {\bibfnamefont {J.~A.}\ \bibnamefont
  {Dix}}\ and\ \bibinfo {author} {\bibfnamefont {A.}~\bibnamefont {Verkman}},\
  }\href {https://doi.org/10.1146/annurev.biophys.37.032807.125824} {\bibfield
  {journal} {\bibinfo  {journal} {Annu. Rev. Biophys.}\ }\textbf {\bibinfo
  {volume} {37}},\ \bibinfo {pages} {247} (\bibinfo {year} {2008})}\BibitemShut
  {NoStop}%
\bibitem [{\citenamefont {Pal}, \citenamefont {Kostinski},\ and\ \citenamefont
  {Reuveni}(2022)}]{pal2018}%
  \BibitemOpen
  \bibfield  {author} {\bibinfo {author} {\bibfnamefont {A.}~\bibnamefont
  {Pal}}, \bibinfo {author} {\bibfnamefont {S.}~\bibnamefont {Kostinski}},\
  and\ \bibinfo {author} {\bibfnamefont {S.}~\bibnamefont {Reuveni}},\ }\href
  {https://doi.org/10.1088/1751-8121/ac3cdf} {\bibfield  {journal} {\bibinfo
  {journal} {J. Phys. A: Math. Theor.}\ }\textbf {\bibinfo {volume} {55}},\
  \bibinfo {pages} {021001} (\bibinfo {year} {2022})}\BibitemShut {NoStop}%
\bibitem [{\citenamefont {Dechant}(2019)}]{dechant2018}%
  \BibitemOpen
  \bibfield  {author} {\bibinfo {author} {\bibfnamefont {A.}~\bibnamefont
  {Dechant}},\ }\href {https://doi.org/10.1088/1751-8121/aaf3ff} {\bibfield
  {journal} {\bibinfo  {journal} {J. Phys. A}\ }\textbf {\bibinfo {volume}
  {52}},\ \bibinfo {pages} {035001} (\bibinfo {year} {2019})}\BibitemShut
  {NoStop}%
\bibitem [{\citenamefont {Hasegawa}\ and\ \citenamefont
  {Van~Vu}(2019)}]{hasegawa2019}%
  \BibitemOpen
  \bibfield  {author} {\bibinfo {author} {\bibfnamefont {Y.}~\bibnamefont
  {Hasegawa}}\ and\ \bibinfo {author} {\bibfnamefont {T.}~\bibnamefont
  {Van~Vu}},\ }\href {https://doi.org/10.1103/PhysRevE.99.062126} {\bibfield
  {journal} {\bibinfo  {journal} {Phys. Rev. E}\ }\textbf {\bibinfo {volume}
  {99}},\ \bibinfo {pages} {062126} (\bibinfo {year} {2019})}\BibitemShut
  {NoStop}%
\bibitem [{\citenamefont {Dechant}\ and\ \citenamefont
  {Sasa}(2020)}]{dechant2020}%
  \BibitemOpen
  \bibfield  {author} {\bibinfo {author} {\bibfnamefont {A.}~\bibnamefont
  {Dechant}}\ and\ \bibinfo {author} {\bibfnamefont {S.-i.}\ \bibnamefont
  {Sasa}},\ }\href {https://doi.org/10.1073/pnas.1918386117} {\bibfield
  {journal} {\bibinfo  {journal} {Proc. Natl. Acad. Sci.}\ }\textbf {\bibinfo
  {volume} {117}},\ \bibinfo {pages} {6430} (\bibinfo {year}
  {2020})}\BibitemShut {NoStop}%
\bibitem [{\citenamefont {Proesmans}\ and\ \citenamefont {van~den
  Broeck}(2017)}]{proesmans2017}%
  \BibitemOpen
  \bibfield  {author} {\bibinfo {author} {\bibfnamefont {K.}~\bibnamefont
  {Proesmans}}\ and\ \bibinfo {author} {\bibfnamefont {C.}~\bibnamefont
  {van~den Broeck}},\ }\href {https://doi.org/10.1209/0295-5075/119/20001}
  {\bibfield  {journal} {\bibinfo  {journal} {EPL}\ }\textbf {\bibinfo {volume}
  {119}},\ \bibinfo {pages} {20001} (\bibinfo {year} {2017})}\BibitemShut
  {NoStop}%
\bibitem [{\citenamefont {Shiraishi}(2017)}]{shiraishi2017}%
  \BibitemOpen
  \bibfield  {author} {\bibinfo {author} {\bibfnamefont {N.}~\bibnamefont
  {Shiraishi}},\ }\href {https://arxiv.org/abs/1706.00892} {\bibfield
  {journal} {\bibinfo  {journal} {arXiv:1706.00892 [cond-mat.stat-mech]}\ }
  (\bibinfo {year} {2017})}\BibitemShut {NoStop}%
\bibitem [{\citenamefont {David}\ and\ \citenamefont
  {Larry}(1987)}]{david1987}%
  \BibitemOpen
  \bibfield  {author} {\bibinfo {author} {\bibfnamefont {A.}~\bibnamefont
  {David}}\ and\ \bibinfo {author} {\bibfnamefont {S.}~\bibnamefont {Larry}},\
  }\href {https://doi.org/10.1080/15326348708807067} {\bibfield  {journal}
  {\bibinfo  {journal} {Commun. Stat. Stoch. Models}\ }\textbf {\bibinfo
  {volume} {3}},\ \bibinfo {pages} {467} (\bibinfo {year} {1987})}\BibitemShut
  {NoStop}%
\bibitem [{\citenamefont {Gillespie}(1977)}]{Gillespie_1977}%
  \BibitemOpen
  \bibfield  {author} {\bibinfo {author} {\bibfnamefont {D.~T.}\ \bibnamefont
  {Gillespie}},\ }\href {https://doi.org/10.1021/j100540a008} {\bibfield
  {journal} {\bibinfo  {journal} {J. Phys. Chem.}\ }\textbf {\bibinfo {volume}
  {81}},\ \bibinfo {pages} {2340} (\bibinfo {year} {1977})}\BibitemShut
  {NoStop}%
\bibitem [{\citenamefont {Koyuk}\ and\ \citenamefont
  {Seifert}(2020)}]{koyuk2020}%
  \BibitemOpen
  \bibfield  {author} {\bibinfo {author} {\bibfnamefont {T.}~\bibnamefont
  {Koyuk}}\ and\ \bibinfo {author} {\bibfnamefont {U.}~\bibnamefont
  {Seifert}},\ }\href {https://doi.org/10.1103/PhysRevLett.125.260604}
  {\bibfield  {journal} {\bibinfo  {journal} {Phys. Rev. Lett.}\ }\textbf
  {\bibinfo {volume} {125}},\ \bibinfo {pages} {260604} (\bibinfo {year}
  {2020})}\BibitemShut {NoStop}%
\bibitem [{\citenamefont {Pietzonka}, \citenamefont {Barato},\ and\
  \citenamefont {Seifert}(2016)}]{pietzonka2016}%
  \BibitemOpen
  \bibfield  {author} {\bibinfo {author} {\bibfnamefont {P.}~\bibnamefont
  {Pietzonka}}, \bibinfo {author} {\bibfnamefont {A.~C.}\ \bibnamefont
  {Barato}},\ and\ \bibinfo {author} {\bibfnamefont {U.}~\bibnamefont
  {Seifert}},\ }\href {https://doi.org/10.1088/1751-8113/49/34/34LT01}
  {\bibfield  {journal} {\bibinfo  {journal} {J. Phys. A: Math. Theor.}\
  }\textbf {\bibinfo {volume} {49}},\ \bibinfo {pages} {34LT01} (\bibinfo
  {year} {2016})}\BibitemShut {NoStop}%
\bibitem [{\citenamefont {Hartich}\ and\ \citenamefont
  {Godec}(2021)}]{hartich2021}%
  \BibitemOpen
  \bibfield  {author} {\bibinfo {author} {\bibfnamefont {D.}~\bibnamefont
  {Hartich}}\ and\ \bibinfo {author} {\bibfnamefont {A.}~\bibnamefont
  {Godec}},\ }\href {https://doi.org/10.1103/PhysRevX.11.041047} {\bibfield
  {journal} {\bibinfo  {journal} {Phys. Rev. X}\ }\textbf {\bibinfo {volume}
  {11}},\ \bibinfo {pages} {041047} (\bibinfo {year} {2021})}\BibitemShut
  {NoStop}%
\bibitem [{\citenamefont {Shiraishi}(2021)}]{shiraishi2021}%
  \BibitemOpen
  \bibfield  {author} {\bibinfo {author} {\bibfnamefont {N.}~\bibnamefont
  {Shiraishi}},\ }\href {https://doi.org/10.1007/s10955-021-02829-8} {\bibfield
   {journal} {\bibinfo  {journal} {J. Stat. Phys.}\ }\textbf {\bibinfo {volume}
  {185}} (\bibinfo {year} {2021}),\ 10.1007/s10955-021-02829-8}\BibitemShut
  {NoStop}%
\bibitem [{\citenamefont {Polettini}, \citenamefont {Falasco},\ and\
  \citenamefont {Esposito}(2021)}]{polettini2021}%
  \BibitemOpen
  \bibfield  {author} {\bibinfo {author} {\bibfnamefont {M.}~\bibnamefont
  {Polettini}}, \bibinfo {author} {\bibfnamefont {G.}~\bibnamefont {Falasco}},\
  and\ \bibinfo {author} {\bibfnamefont {M.}~\bibnamefont {Esposito}},\ }\href
  {https://arxiv.org/abs/2106.00425} {\bibfield  {journal} {\bibinfo  {journal}
  {arXiv:2106.00425 [cond-mat.stat-mech]}\ } (\bibinfo {year}
  {2021})}\BibitemShut {NoStop}%
\bibitem [{\citenamefont {Schnakenberg}(1976)}]{schnakenberg1976}%
  \BibitemOpen
  \bibfield  {author} {\bibinfo {author} {\bibfnamefont {J.}~\bibnamefont
  {Schnakenberg}},\ }\href {https://doi.org/10.1103/RevModPhys.48.571}
  {\bibfield  {journal} {\bibinfo  {journal} {Rev. Mod. Phys.}\ }\textbf
  {\bibinfo {volume} {48}},\ \bibinfo {pages} {571} (\bibinfo {year}
  {1976})}\BibitemShut {NoStop}%
\bibitem [{\citenamefont {Andrieux}\ and\ \citenamefont
  {Gaspard}(2007)}]{andrieux2007}%
  \BibitemOpen
  \bibfield  {author} {\bibinfo {author} {\bibfnamefont {D.}~\bibnamefont
  {Andrieux}}\ and\ \bibinfo {author} {\bibfnamefont {P.}~\bibnamefont
  {Gaspard}},\ }\href {https://doi.org/10.1007/s10955-006-9233-5} {\bibfield
  {journal} {\bibinfo  {journal} {J. Stat. Phys.}\ }\textbf {\bibinfo {volume}
  {127}},\ \bibinfo {pages} {107} (\bibinfo {year} {2007})}\BibitemShut
  {NoStop}%
\bibitem [{\citenamefont {Puglisi}\ \emph {et~al.}(2010)\citenamefont
  {Puglisi}, \citenamefont {Pigolotti}, \citenamefont {Rondoni},\ and\
  \citenamefont {Vulpiani}}]{puglisi2010}%
  \BibitemOpen
  \bibfield  {author} {\bibinfo {author} {\bibfnamefont {A.}~\bibnamefont
  {Puglisi}}, \bibinfo {author} {\bibfnamefont {S.}~\bibnamefont {Pigolotti}},
  \bibinfo {author} {\bibfnamefont {L.}~\bibnamefont {Rondoni}},\ and\ \bibinfo
  {author} {\bibfnamefont {A.}~\bibnamefont {Vulpiani}},\ }\href
  {https://doi.org/10.1088/1742-5468/2010/05/P05015} {\bibfield  {journal}
  {\bibinfo  {journal} {J. Stat. Mech.}\ }\textbf {\bibinfo {volume} {2010}},\
  \bibinfo {pages} {P05015} (\bibinfo {year} {2010})}\BibitemShut {NoStop}%
\end{thebibliography}
\bibliographystyle{aipnum4-2}

\end{document}